\documentclass{article}

\immediate\write18{sh ./vc}
\gdef\GITAbrHash{1d3b604}%
\gdef\VCRevision{\GITAbrHash}%
\gdef\VCDateISO{2013-05-16}%
\usepackage[utf8]{inputenc}
\usepackage[T1]{fontenc}
\usepackage{lmodern}
\usepackage{tikz}
\usepackage{subfig}
\usepackage{nicefrac}
\usepackage[squaren]{SIunits}
\usetikzlibrary{positioning}
\usepackage{caption}
\usepackage{lipsum}
\usepackage{textcomp}
\usepackage{graphicx}
\usepackage{booktabs}
\usepackage{amsmath, amssymb}
\usepackage[
a4paper, bookmarks, breaklinks,
     pdfauthor={Giovanni Luca Ciampaglia}
]{hyperref}  

\newcommand{\nf}[2]{\nicefrac[\textrm]{#1}{#2}}

\usepackage{fancyhdr}
\usepackage{xcolor}
\let\mymarginpar\marginpar
\renewcommand{\marginpar}[3]{%
\mymarginpar{\raggedright\hbadness=10000\footnotesize\it #1\par}}

\title{A framework for the calibration of social simulation models}

\author{Giovanni Luca Ciampaglia\\
    \multicolumn{1}{p{.7\textwidth}}{\small\centering\emph{
    Center for Complex Networks and Systems Research,
    School of Informatics and Computing, Indiana University\\
    919 E 10th St, Bloomington, IN 47408, USA\\
    }gciampag@indiana.edu}
}

\begin{document}

\maketitle

\thispagestyle{fancy}

\begin{abstract} Simulation with agent-based models is increasingly used in the
    study of complex socio-technical systems and in social simulation in
    general. This paradigm offers a number of attractive features, namely the
    possibility of modeling emergent phenomena within large populations. As a
    consequence, often the quantity in need of calibration may be a distribution over the
    population whose relation with the parameters of the model is analytically
    intractable. Nevertheless, we can simulate. In this paper we present a
    simulation-based framework for the calibration of agent-based models with
    distributional output based on indirect inference. We illustrate our method
    step by step on a model of norm emergence in an online community of peer
    production, using data from three large Wikipedia communities. Model fit and
    diagnostics are discussed. \end{abstract}

\section{Introduction}

Computational agent-based models (\textsc{abm}) are increasingly used in several
areas of science because of a number of attractive features
\cite{Bonabeau2002a}. They are an effective alternative to more traditional
methods because they let one test \emph{in silico} different hypotheses about
the origin of collective phenomena, and to explore the connection
between the micro and macro levels, i.e. what kind of macroscopic patterns are
generated starting from a set of microscopic interactions between agents.
\cite{Epstein1996a}. 

Like most models, \textsc{abm}s contain tunable parameters and thus, as a first
step towards empirical investigation, it is desirable to calibrate them.
However, often it is not possible to formulate an equivalent analytical model,
and even if this is the case, it may still be intractable or too complicated,
for example when latent variables are used. In all these cases, an \textsc{abm}
can be just regarded as a black box, and be calibrated via simulation, using the
machinery developed for complex computer codes \cite{Santner2003a}. These
approaches typically involve the use of emulators such as splines, polynomials,
or semi-parametric techniques such as Gaussian Processes \cite{Santner2003a}.
Other approaches to calibration that are not simulation-based but still use
Gaussian processes are Bayesian techniques
\cite{Kennedy2001,Higdon2005,Bayarri2007}; they have been applied in biology
\cite{Dancik2010a} and cosmology \cite{Heitmann2006}, to cite a few.
Computational approaches for the calibration of \textsc{abm}s have been
developed in biology \cite{Ellner2006a,Dancik2010a} and economics
\cite{Bianchi2007a,Gilli2003a,Windrum2007}. Comparatively, issues related to
calibration and empirical testing are still largely underrepresented in the
social-simulation literature \cite{Sobkowicz2009a}.

Agent-based models, though, present another difficulty to the above approaches:
it is often the case that the output of an \textsc{abm} is a distribution of
values over a population of agents, and not just a scalar or vector-valued
quantity. This is common, for example, in models of social collective phenomena.
Sometimes distributions can be summarized by their moments and thus methods like
the Simulated Method of Moments may be applied \cite{McFadden1989}. Some other
times the distribution moments do not provide a good description of the output
distribution, and other approaches must be used
\cite{Bianchi2007a,Dancik2010a,Ellner2006a}.

In this paper we present a step-by-step data-driven methodology for calibrating
an agent-based model with distributional output. We use a computational
technique inspired by the indirect inference methodology \cite{Gourieroux1993a}.
This technique predicates the use of an \emph{auxiliary} model to match
empirical data with synthetic simulations. 

Indirect inference is used to fit models to empirical data when maximum
likelihood estimation is either unfeasible, or simply too complicated from a
computational point of view. Examples from econometrics -- where the technique
first originated -- include dynamical models with latent variables (cf.
\cite{Gourieroux1993a,Smith1993}), agent-based models \cite{Bianchi2007a}, and
also dynamic models from population biology \cite{Kendall2005a,Wood2010a}. A
different, related approach based on simulation is the one by Gilli and Winker
\cite{Gilli2003a}. 

We apply our technique to the calibration of a model of a techno-social system,
specifically a model of norm formation in an online community of social
production or commons-based peer production (shortened as ``peer production''
in the remainder of the text) \cite{Ciampaglia2011}. 

Norms are shared expectations about behaviors that members of a social group
ought follow or else incur in the risk of being sanctioned by other members
\cite{Opp2001}. Norms are regarded as one of the main determinants of social
behavior, a sort of grammar for social interactions \cite{Bicchieri2005}, and
are central to the understanding of group dynamics \cite{Feldman1984}. The
emergence of social norms has been the subject of a long-standing tradition of
investigation \cite{Hamilton1964,Hardin1982,Ostrom2000} and, more recently, also
by means of agent-based models \cite{Axelrod1986,Bicchieri2005,Alexander2007}.
Norms may emerge from imposition from a higher authority \cite{Opp2001}, but
often -- and this is the case we are interested in with this paper -- norms can
emerge informally from the aggregated behavior of many different actors
\cite{Axelrod1986,Bicchieri2005}. 

Social norms are important determinants of behavior in online peer production
groups, where members engage, often for free \cite{Lerner2002a}, in the
production of digital contents \cite{Benkler2006}. In these settings users may
be encouraged to contribute to the digital common by a variety of extrinsic
rewards, from reputation gains \cite{Ciffolilli2003} to simple forms of
acknowledgment \cite{Cheshire2008}. Furthermore, direct surveys show that the
array of intrinsic motivations for contributing is surprisingly varied
\cite{Rafaeli2008a}. 

However, in many cases neither rewards nor personal motivations can foster true,
long-term commitment to the community without the construction of a shared sense
of membership \cite{Ren2012}. This can be construed in terms of social identity
\cite{Tajfel1982} and self-categorization \cite{Turner1989}. As a result, a
strong, shared identity may develop, even despite the fact that interactions on
the Web are often asynchronous and anonymous. For example, recent work by Neff
et al. analyzed the discourse of editors from the English Wikipedia and
discovered that the `Wikipedian' identity can be stronger than affiliation to
the two major parties of the US political system \cite{Neff2012}.

A peculiarity of peer-production groups is that norms may also specifically
regulate how, and under which conditions, users contribute to the digital
common. An example of a social production norm is the Neutral Point of View
(\textsc{npov}) policy of Wikipedia. This policy prescribes users ``\emph{to
provide complete information, and not to promote one particular point of view
over another}'' when contributing to encyclopedic articles.\footnote{See
\url{http://goo.gl/Jw8Ic}.} Users who do not frame their contributions following
the \textsc{npov} guidelines are sanctioned by other peers by having their
contributions rejected. Thus, besides the low barriers to contribution
\cite{Ciffolilli2003}, the emergence of the proper, efficient social production
norms are fundamental to the success of an online community of peer production
as much as for any social group \cite{Feldman1984}. An example of this is the
establishment of a ``good-faith'' collaboration culture in Wikipedia given by
Reagle \cite{Reagle2007}.

On the other hand, certain peculiar characteristics of online peer-production
groups pose challenges to the study of the emergence of social norms. Mass
collaboration platforms akin to Wikipedia may reach considerable sizes,
and the churn among users can become surprisingly large, meaning that previous
research, predicated on small group sizes and consistent participation
\cite{Graham2003}, is difficult to apply.\footnote{As of December 2012, there
    were 796,945 registered editors with at least ten contributions in the
    English Wikipedia; the peak of activity was in March 2007, when more than
    51,000 `Wikipedians' (registered editors who made at least ten edits)
    contributed five or more edits in that month. Of those, nearly 14,000 had
    registered that same month. See \url{http://stats.wikimedia.org}.} 

Simple microscopic models of collective behavior are able to account for
striking regularities in social phenomena, a classic example being the case of
elections \cite{Fortunato2007a}. The model we propose in this paper addresses all
the above problems by embedding a process of belief adjustment based on
homophily \cite{McPherson2001a} within a dynamic population
\cite{Deffuant2001a,Hegselmann2002a}. Social production norms emerge informally
from the microscopic interactions between agents and the digital common, which
in our case will be a set of pages, or artifacts. Thus our model falls within
the category of norm-emergence models with structured interactions
\cite{Alexander2007}. 

Unlike the other models of norm emergence cited so far, linking the process of
norm emergence with the population dynamic gives the possibility to study the
emergence of norms in a more realistic setting and, as a collateral benefit, to
test it against empirical data about user participation, such as the span of
user activity. We use indirect inference and our model to analyze data on user
activity from three large Wikipedia communities (French, Italian, and
Portuguese). 

Because our objective for this paper is mainly to illustrate an \textsc{abm}
calibration methodology, it would have seemed more sensible to set aside the
problem of norm emergence in peer production groups and to focus instead on a
well-established model from the literature on social simulation, such as the
Schelling segregation model \cite{Schelling1971} or the Axelrod model of
cultural evolution \cite{Axelrod1997a}, to cite a few. Classic models have the
advantage of a clear, extensively studied phenomenology, and usually possess few
parameters, making them an effective choice for illustrative purposes. On the
other hand, they are often very idealized and sometimes fail at reproducing
simple empirical evidence.\footnote{In the original model of Axelrod, for
    example, cultural diversity is more likely to occur in small groups rather
    than large groups, contrary to basic empirical evidence; cf. the work by
    Flache and Macy on this and other shortcomings of the Axelrod model
    \cite{Flache2011a}. The model of Schelling has often been put to empirical
    test using data from surveys, but usually simulation requires to introduce a
    number of parameters that is comparable to the one of the model we use
    \cite{Clark2008}, thus defeating its original advantage in terms of
simplicity.}

Finally, the choice of studying norm emergence in an online setting can be
further motivated considering that studying norm emergence in an online setting
addresses a classic problem with modeling social norms and human behavior in
general: while data about preferences of social actors can be collected via
experimentation both online and offline \cite{Bicchieri2005,Camerer2004},
large-scale online groups still present several problems with respect to this
task. In contrast, data about social interactions in said groups are nowadays
comparatively much easier to obtain, as abundant traces of human activities are
readily available on the Internet \cite{Lazer2009}.

The model we use here is a modification of the original model by
\cite{Ciampaglia2011}, that employed homogeneous Poisson processes to
describe the patterns of temporal activation of agents -- an assumption that is
not in line with recent findings on human activity on the Internet
\cite{Radicchi2009a}. We found that this assumption has a profound impact on the
resulting distribution of user lifespans, which makes the model not amenable for
calibration against empirical data. We thus decided to lift the homogeneity
assumption and introduce a realistic sub-model of temporal activation of users
based on Poissonian cascades \cite{Malmgren2008a}.

The rest of the paper is organized as follows: in section \ref{s:data} we
describe the collection and preparation of the user activity span data; in
section \ref{s:model} we describe the model of norm emergence in a community of
peer production; section \ref{s:methods} illustrates the calibration technique
in detail; section \ref{s:results} gives the results of the calibration,
including various diagnostics procedures. We critically evaluate the results of
the calibration and conclude in section \ref{s:discussion} and finally conclude
(sec. \ref{s:conclusions}).

\section{Data} 
\label{s:data}

We measure the period of participation of a user within an online group as the
span between the first and the last contribution she makes. We call this measure
the user activity lifespan $\tau$. Previous research on peer-production systems
shows that the period of activity of users in blogs, wikis, etc. follows a
multimodal distribution \cite{Guo2009a,Ciampaglia2010a,Wu2010,Zhang2012}. Data
on user activity lifespans of Wikipedia users were obtained from the official
database dumps released by the Wikimedia Foundation. We used the 2009
\texttt{stub-meta-history} dumps, which contain only the metadata of the
revisions.\footnote{See \url{http://dumps.wikimedia.org}.} 

We selected the top five languages in terms of authored articles, which, as of
data collection time, were  English (3,7M articles as of September 2011), German
(1,3M), French (1,1M), Italian (850K), and Portuguese (700K). In terms of
activity lifespan they represent roughly two distinct classes, depending on the
ratio between short-term and long-term users. One class, which comprises
Portuguese and English, has a more even ratio than the other (French, Italian
and German) \cite{Ciampaglia2010a}. The question of whether these two classes of
user activity lifespan are representative of the whole catalog of Wikipedia
communities is still an open one.

While we initially collected data from these five different Wikipedia
communities, computational issues with the English and the German wikis
eventually forced us to drop these two, which were the most populated and thus
could not be simulated in a reasonable amount of time (i.e., they required $> 1$
month), given our limited resources. Because each community is simulated
independently of the others, exclusion of these two dataset does not impact the
results about the other three. It also leaves us with at least one specimen per
activity lifespan class. We thus decided to report results only from the
remaining three: French, Italian, and Portuguese. For a histogram of the
lifespan data, see Figure \ref{fig:fitresults}.

The raw data from the dumps includes details about each revision to a Wikipedia
article, the time stamp, user \textsc{id},  user name, and additional
information. As an example of the raw data format, Table \ref{tab:data} reports
five consecutive revisions taken from a page of the Italian Wikipedia. Anonymous
contributions have all \textsc{id} $= 0$ and the \textsc{ip} address of the
originating computer instead of the user name. 

We discarded all anonymous contributions and all revisions of non-human users
(i.e. bots),\footnote{To do this, we cross referenced data from the
\texttt{user\_groups} table, which is also available at the dumps website.}
grouped revisions by the user \textsc{id}, and sorted them chronologically, so
that for each user $u$ we obtained revision times $t_1^{\left( i \right)} <
\ldots < t^{\left( i \right)}_{N_i}$, where $N_i$ is the total number of edits
of the $i$-th user. The lifespan of the $i$-th user is thus $\tau_i = t^{\left(
i\right)}_{N_i} - t^{\left( i \right)}_1$. 

\begin{table}
    \footnotesize
    \centering
    \caption{Raw data excerpt from the Wikipedia article about Pope Clement VII
        from the Italian Wikipedia. See \url{http://goo.gl/zCB8m} for the online
    version.}
    \begin{tabular}{@{\extracolsep{\fill}}rlc@{\extracolsep{0pt}}}
        \toprule
        user \textsc{id} & user name & revision time stamp    \\
        \midrule
        7077 & Moroboshi     & 2006-05-26 04:37:45 \\
        0 & 82.50.4.229   & 2006-05-26 19:15:38 \\
        36426 & Sailko        & 2006-06-05 08:32:48 \\
        57872 & Dapa19        & 2006-06-07 16:31:58 \\
        35813 & Moloch981     & 2006-06-07 20:24:14 \\
        \bottomrule
    \end{tabular}
    \label{tab:data}
\end{table}

It should be noted that $\tau$ is a proxy for the true period of participation
of a user. In our model, we have access to the latter, so care must be taken
when comparing the model output with the empirical data. This is further
discussed in Section \ref{ss:patterns} when describing the actual output of the
model.

\section{A model of emergence of social production norms}
\label{s:model}

In this section we describe the agent-based model used in the paper. The model
is comprised of different mechanisms that describe when agents perform actions,
how they interact with other agents, and how these interactions lead to a change
in the state of the agents. 

\subsection{Norm emergence}

We model the emergence of social production norms as a process of adjustment of
shared beliefs about how people ought to contribute to the digital common. For
example, we can consider whether Wikipedia users will adopt a neutral style of
writing or not. The concept of a neutral point of view is of course nuanced and
multifaceted and thus it seems plausible to model it as a quantity ranging
within a spectrum of possibilities, instead of just dichotomously. That is,
instead of saying that a user either writes neutrally or non-neutrally (e.g. $x$
is either 0 or 1) we could say that there is a range of possible alternatives
(for simplicity we consider only one dimension) and that, of all these
alternatives, there might be within the group a shared value that is socially
acceptable, i.e., a consensus. In this case we say that a norm has emerged.

Thus we employ a continuous framework similar to the models of opinion formation
by Deffuant et al. \cite{Deffuant2001a} and by Hegselmann and Krause
\cite{Hegselmann2002a}, which explain under which conditions a group of people
will, through repeated interactions, reach a consensus about a set of beliefs,
values, or opinions. In these models people adjust their opinions based on who
they interact with. The empirical assumption is that belief adjustment is based
on homophily; that is, people will effectively adjust their opinion only if they
interact with other people that are sufficiently similar to them. 

To adapt this framework to the context of norms and behavior in a peer
production group, we need to change the interpretation we give to these
quantities. In our model there are two types of agents: \emph{users} and
\emph{pages}. Each user is represented by a real dynamic variable $x_i\left( t
\right)\in \left[ 0, 1 \right]$. Similarly, the $j$-th page is described by real
variable $y_j\left( t \right)\in \left[ 0,1 \right]$. 

We can imagine that the following events may occur in our model: first, a user
$i$ contributes to pages according to what she believes is socially appropriate
behavior (i.e. her belief $x_i$). Moreover, she may change her beliefs about
what is appropriate by interacting with a page $j$ that shows how other users
contributed to the common (i.e. the page state $y_j$), provided that these are
sufficiently similar to hers (i.e. $\left|x_i - y_j\right| < \varepsilon$), thus
employing the empirical pattern of homophily. In the language of the models of
opinion formation, this is called \emph{bounded confidence} (\textsc{bc}). In
particular, we use the bounded confidence rule from the model by Deffuant et al.
\cite{Deffuant2001a}: let $t$ be time the $i$-th user interacts with the $j$-th
page. If $\left|x_i\left( t \right) - y_j\left( t \right)\right| < \varepsilon$
then,
\begin{eqnarray} x_i\left( t \right) &\gets& x_i\left( t \right) + \mu\left(
    y_j\left( t \right) - x_i\left( t \right)\right) \label{eq:mod:1} \\
    y_j\left( t \right) &\gets& y_j\left( t \right) + \mu\left( x_i\left( t
    \right) - y_j\left( t \right)\right) \label{eq:mod:2} \end{eqnarray}

The \emph{speed} or \emph{uncertainty} parameter $\mu\in\left( 0,\nf{1}{2}
\right]$ governs the entity of the belief adjustment. The parameter
$\varepsilon\ge 0$ dictates the range of social influence and is called the
\emph{confidence}. 

The interpretation often given to extremal values of the opinion space in
classic models of opinion dynamics is that of `extreme' opinions on a specific
topic of discussion. For example the $\left[\,0, 1 \right]$ space might
represent the political spectrum and the values 0 and 1 might represent the
points of view of the extreme Left and extreme Right. Under this interpretation
it is thus interesting to see what happens if one assumes that the confidence
$\varepsilon$ of an agent depends on its opinion value, in order to mimic the
empirical observation that extreme opinions are often accompanied by narrow
homophilistic bounds of confidence \cite{Deffuant2002a,Hegselmann2002a}. This
assumption finds support in the pioneering work by Moscovici et al.
\cite{Moscovici1969} on minority influence on a perceptual task.

Since our interpretation of the dynamic quantities $x$ and $y$ is different from
opinion formation, it is not clear what it would mean to have asymmetric
confidence bounds in this framework. That is, in our framework a consensus near
the bounds of the belief domain does not imply that an `extremist' norm has
emerged, only that consensus has taken place there. This obviously does not mean
that heterogeneity in the population might not have interesting effects on the
dynamic of belief adjustment, as Hegselmann and Krause showed
\cite{Hegselmann2002a}, but, for simplicity, in this model we assume that the
confidence bounds are the same over the whole population of agents; that is,
there is no asymmetry.

Of course, norms differ from pure informal social conventions that emerge, for
example, from coordination, because not abiding to a norm may result in being
sanctioned \cite{Bicchieri2005}. In our case it is plausible to think that
doing a modification to a page for sanctioning purposes does not involve a
change in the beliefs of who is sanctioning, but only in those shown by the
page, e.g.  Wikipedia editors correcting vandalism. Of course,
sanctioning involves a cost, and thus it does not happen all the time. In some
cases, reverting the work of others may be cumbersome, in other cases not.
The possibility to easily revert vandalism is credited as one of the main
reasons for the success of Wikipedia \cite{Ciffolilli2003}. 

To account for this in our model, we introduce another possibility when a user
interacts with a page. If $\left|x_i\left( t \right) - y_j\left( t
\right)\right| \ge \varepsilon$, then with probability $p_{\mathrm{rollback}}$ only eq.
\eqref{eq:mod:2} applies. This is intended to model the act of reverting the
work of the preceding editor, as described above. That is, even though
similarity, because of the homophily assumption, prompts a modification to both
the user belief $x_i$ and the feature of the page $y_i$, dissimilarity may, with
probability $p_{\mathrm{rollback}}$, still cause a modification of the page, thus allowing for an
indirect sanctioning mechanism of the people who edited the page before. We call
the parameter $p_{\mathrm{rollback}}$ the \emph{rollback probability} as a reference to the rollback
(or revert) feature of Wikipedia, which allows editors to restore the current
version of a page to a previous revision. This mechanism is reminiscent of
interaction noise of models of opinion dynamics \cite{Maes2010}. 

\subsection{User activity lifespan}

Users contribute to the digital common of the community and are active in the
community only for a certain period, the \emph{activity lifespan} $\tau$. The
population of users is open: at every time new users join and existing
users retire. The rate at which new users join will depend on different external
factors, such as the popularity of the project and the barriers to contribution.
For simplicity, we consider that new users join at a fixed rate $\rho_u$. 

The rate at which existing users retire, instead, will plausibly depend on the
incentives and motivations of the users \cite{Deci1999}. Research on retention
of new members on Wikipedia has shown in fact that the rejection of
contributions is especially demotivating for newcomers \cite{Halfaker2011}. For
more tenured editors, on the other hand, the perceived quality of the community
as a whole \cite{Hirschman1970} could be an important factor. Finally, people
cannot assess factors like these accurately and are instead more likely to
resort to simple heuristics \cite{Hardin1982,Alexander2007}. 

We would like to have a single quantity that summarizes all these
considerations. Let us denote with $n_i\left( t \right)$ the total number of
contributions a user made at time $t$,  with $s_i\left( t \right)$ the number of
times in which Eq. \eqref{eq:mod:1} was applied, that is, the number of times in
which she was `influenced' by the page. A simple way to model how user $i$
perceives her `success' within the community could thus be:

\begin{equation} r_i\left( t \right) = \frac{s_i\left( t \right) + c_s}{n_i\left(
    t \right) + c_s} \label{eq:mod:4} \end{equation}

\noindent where $c_s \ge 0$ is a constant that represents the initial motivation
users have when joining the community. The higher $c_s$ is, the larger the
number of rejections needed to induce them to retire. As the number of
contributions $n_i\left( t \right)$ grows, the effect of $c_s$ will become
smaller, and the overall commitment $r_i$ of a user will be determined by the
perceived quality of the project, as estimated from the sample of pages she
interacted with. The ratio $\nicefrac{s_i}{n_i}$ can also be regarded as an
estimate that the user has about how much her current belief is in accordance
with the rest of the group \cite{Bicchieri2005}.

Because $r_i$ depends on $s_i$ and $n_i$, its temporal evolution will depend on
the rate at which users contribute to the community. Of course, if her first
contribution is rejected, then chances are the a user might still try more
before giving up completely, which means that even if $r_i = 0$, the expected
short-term activity lifespan will be a value $\tau_1$ that does not depend on
the frequency at which users perform edits. On the other hand, if $r_i = 1$ a
user may still decide to retire for other external factors that may be largely
personal and thus hard to summarize, so it is reasonable to assume that there is
also a natural lifespan of long-term activity $\tau_0$, past which even the most
committed member will stop participating. A simple way to capture this is to
define the rate of departure $\lambda^{\left( i \right)}_d\left( t \right)$ of
user $i$ at time $t$ as:

\begin{equation} \lambda^{\left( i \right)}_d\left( t \right) = \frac{r_i\left( t \right)}{\tau_0} +
    \frac{1 - r_i\left( t \right)}{\tau_1} \label{eq:mod:3}
\end{equation}

where we implicitly also assume $\tau_0 \gg \tau_1$; i.e. they effectively refer
to different time scales. It should be noted that Eq. \eqref{eq:mod:3} simply
interpolates $r_i$ and thus the actual proportion of short-term and long-term
users, i.e. between `infant mortality' and `wear out', will depend
exclusively on the collective dynamics of $r_i$, i.e. Eq. \eqref{eq:mod:3}
simply constrains the location of the distribution of $\tau$, but not its actual
shape.

\subsection{Temporal activation patterns}\label{ss:patterns}

In the original model users interacted with pages at a constant rate $\lambda_e$
\cite{Ciampaglia2011}. However, a homogeneous process is not capable of
capturing one essential aspect of the real editing activity of users in an
online community -- \emph{burstiness}. In a broad range of user activities (e.g.
emails, stock trading, phone calls, \textsc{sms}, etc.) it is common to observe
the existence of clusters of events, that is, events that tend to happen in
rapid sequence, separated by long periods of inactivity \cite{Barabasi2005a}.

We found that the seemingly innocuous assumption of homogeneous activity has
strong implications on the distribution of user lifespans produced by the
original model. As we described in Section \ref{s:data}, the user lifespan
$\tau$ is the period between the first and the last contributions of a user.
This obviously means that we do not have data for those users who have less than
two edits. To make the comparison possible, then, we decided to code the model
so that it would output only the sequence of edits performed, and not the
simulated lifespans. 

Let us now consider a new user whose first interaction occurs with a page s.t.
$\left|x_i - y_i\right| \ge \varepsilon$. As we described in Section
\ref{s:model}, this counts as an `unsuccessful' interaction and thus, after
this $s_i = 0$ and $n_i = 1$. Let us also assume, for simplicity, that $c_s =
0$. Then, according to Eq. \eqref{eq:mod:4}, this user will have $r_i\left( t
\right) = 0$ and, from Eq. \eqref{eq:mod:3}, an expected average survival time of
$E\left[ \tau \right] = \tau_1$, where $\tau_1$ is the short-term time scale
parameter. Now, if the mean interval between two consecutive edits $\tau_e =
\lambda_e^{-1}$ is considerably higher than $\tau_1$, the user will become
inactive before performing the second edit, and thus she will not be included in
the output of the model. Thus, assuming homogeneous editing activity
introduces an artificial truncation of the data in that we do not see, on
average, lifespan observations $\tau$ below $\lambda_e$. 

To overcome the above problem, we lift the homogeneity assumption and instead
consider a model of cascading editing events \cite{Malmgren2008a}. As before, we
assume that a user edits pages with a constant rate $\lambda_a$. Once she is
active, she performs, on average, $N_a$ additional edits with rate $\lambda_e$
-- a \emph{cascade} of edits. This has the effect of decoupling the editing rate
from the short-term lifespan, provided of course that $\tau_0$ is also greater
than $\lambda_a^{-1}$. In practice, we set it s.t. $\lambda_a^{-1} \approx
\frac{\tau_0}{2}$. A bursty pattern of activity is obtained by assuming
$\lambda_e \gg \lambda_a$, i.e. the rate of editing within an session is larger
than the rate of editing sessions.

\subsection{Page creation and selection}

Finally, we need to specify how pages are created and modified. We may imagine
that new pages are continuously added to the project. Again, the rate at which
this will happen will depend on a host of external factors, and thus we can
simply assume that pages are created at a constant rate $\rho_p$. Pages are
selected for editing according to preferential attachment \cite{Barabasi1999}.
Once a user activates to perform an edit, a page is chosen with probability
proportional to the number of edits $k$ it has already received, plus a
popularity dampening factor $c_p\ge 0$, which controls how much popular pages, as measured
by the number of contributions received $k$, are more likely to be selected over
non popular pages. In the limit $c_p\rightarrow\infty$, we recover a uniform
distribution.

\subsection{Model implementation}

Let us consider a population of users and a pool of pages. There are four
possible events that we need to consider:

\begin{enumerate}
    \item A new user joins the community.
    \item A new page is created by some user.
    \item A user retires permanently.
    \item A user starts an editing session.
\end{enumerate}

The first two events are homogeneous Poisson processes. The third event is an
inhomogeneous process, and therefore we need to keep track of the rates
$\lambda_d^{\left( i \right)}$ in a dynamic array to which we can add or remove
users.

The last event models the presence of editing cascades. Because the activation
of an editor induces a cascade of edits and not simply a single edit, we need to
keep track of which editors are currently active. We do this by means of a
queue. In practice, whenever an editor activates with rate $\lambda_a$, we put
in the queue $N_a$ edits. When we insert an edit event in the queue we draw the
time of the edit with rate $\lambda_e$. 

The overall population dynamic is simulated using the Gillespie algorithm
\cite{Gillespie1977a}. The main simulation loop is structured as follows. Based
on the number of users and pages, we compute the global rate for each of the
five events and from these the global rate of activity $\lambda$. With this, we
extract the time of the next event. We then check the queue to see if there are
any edits that happen before the next event and perform them, selecting the page
and updating their state. Then we draw the type of the event and, conditional on
this, the information needed to update the state of the system. If an editor
retires, we scan the queue and remove all the remaining edits with her index.
Once the system is updated, we recompute all the global rates for each class of
events, and proceed with the next iteration of the loop.

The model was implemented in Python, with optimizations of the most
computationally expensive parts made in Cython. A run of the model with the
parameterization for the Portuguese Wikipedia (see Table \ref{tab:fixedparams})
took on average 15 minutes on a low-end workstation with 4GB of \textsc{ram}.

\section{Methods} 
\label{s:methods}

\subsection{Overview}

The goal of our estimation technique is to calibrate the peer production model
on the distribution of user activity lifespans of existing Wikipedia
communities. The calibration technique we propose is inspired by the indirect
inference method (see Section \ref{ss:indinf}), but it differs from the classic
indirect inference methodology in a few details. 

Traditional indirect inference assumes that model evaluations are
computationally cheap. On the other hand, depending on the value of the
parameters, simulation of an ordinary agent-based model may take minutes up to
hours. Therefore, in order to apply the framework of indirect inference, we need
to speed up the evaluation phase of the computer model. Our approach is to
combine indirect inference with the use of a surrogate model, based on a
Gaussian process (from now on \textsc{gp}), to approximate the computer model
code.

Application of surrogate models is straightforward when the output of the
computer model is univariate or multivariate but with few variables. With
high-dimensional outputs, for example time series, one can first use a
dimensionality reduction technique on the data. For example, Dancik et al. use
principal component analysis (\textsc{pca}) to reduce a time series output
\cite{Dancik2010a}. 

In our case, direct application of \textsc{gp} is not possible because the
output of the model is a full sample drawn from an unknown, multimodal
distribution of lifespan observations. Therefore, direct application of a \textsc{gp} is not
feasible. We use a Gaussian mixture model (from now on \textsc{gmm}; see Section \ref{ss:gmm}) to perform the pre-processing step.

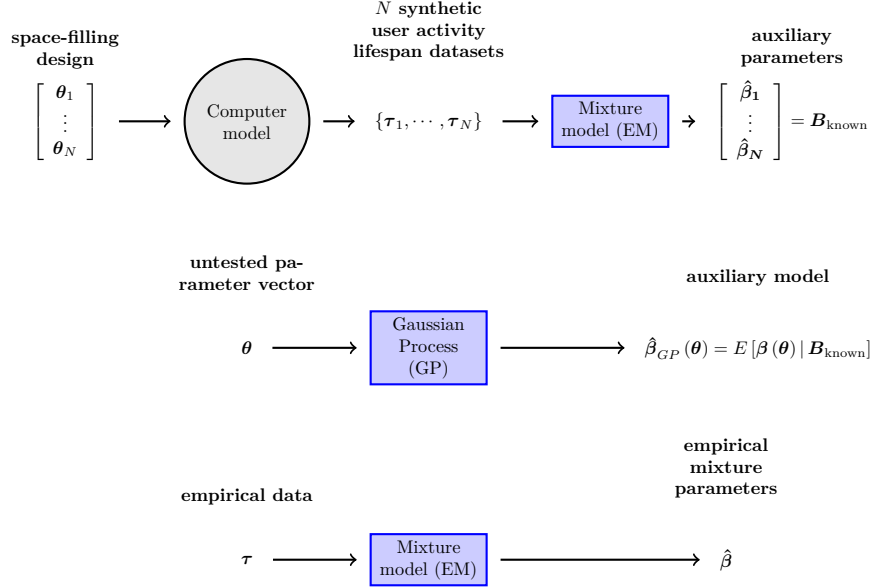
\begin{figure}[t] 
    \begin{center} 
        \begin{tikzpicture}
            [every node/.style={node distance=3.5cm,thick,text centered,scale=.69},
            simulation/.style={circle,draw=black!100,fill=gray!20},
            estimation/.style={rectangle,draw=blue!100,fill=blue!20},
            result/.style={rectangle,draw=black!0}, 
            title/.style={text width=3cm,font=\bf}, 
            post/.style={->,shorten >=5pt, shorten <=5pt, thick}]
            \node [result] (design) {
            $\left[ \begin{array}{c}
                \boldsymbol\theta_1 \\ \vdots \\ \boldsymbol\theta_N \\
            \end{array} \right]$
            }; 
            \node [simulation,text width=2cm] (code) [right of=design] {Computer model};
            \node [result] (result 1) [right of=code] {
            $\left\{ \boldsymbol\tau_1, \cdots, \boldsymbol\tau_N \right\}$
            }; 
            \node [estimation,text width=2cm] (em) [right of=result 1] {Mixture model (EM)}; 
            \node [result] (result 2) [right of=em] {
            $\left[ \begin{array}{c}
                \boldsymbol{\hat\beta_1}\\ \vdots\\ \boldsymbol{\hat\beta_N}\\
            \end{array} \right]= \boldsymbol B_\mathrm{known}$
            }; 
            \node [title,above] at (design.north) {space-filling design}; 
            \node [title,above=.5cm] at (result 1.north) {
            $N$ synthetic user activity lifespan datasets
            }; 
            \node [title,above] at (result 2.north) {auxiliary parameters}; 
            \draw[post] (design) -- (code); 
            \draw[post] (code) -- (result 1); 
            \draw[post] (result 1) -- (em); 
            \draw[post] (em) -- (result 2);
            \node [result] (untested) [below=2cm of code] {$\boldsymbol\theta$}; 
            \node [title,above=.5cm] at (untested.north) {untested parameter vector}; 
            \node [estimation, text width=2cm] (gp) [right of=untested] {Gaussian Process (GP)}; 
            \node [result] (auxiliary) [right=2cm of gp] {
            $\boldsymbol{\hat\beta}_{GP}\left( \boldsymbol\theta \right)= E\left[
            \boldsymbol\beta\left( \boldsymbol\theta \right) |\, \boldsymbol
            B_\mathrm{known}\right]$ 
            }; 
            \node [title,above=.5cm] at (auxiliary.north) {auxiliary model}; 
            \draw[post] (untested) -- (gp);
            \draw[post] (gp) -- (auxiliary);
            \node [result] (data) [below=2.5cm of untested] {$\boldsymbol\tau$};
            \node [title,above=.5cm] at (data.north) {empirical data};
            \node [estimation, text width=2cm] (em 2)  [right of=data] {Mixture model (EM)};
            \node [result] (result 3) [right=3cm of em 2] {$\boldsymbol{\hat\beta}$};
            \node [title,above=.5cm] at (result 3.north) {empirical mixture parameters};
            \draw[post] (data) -- (em 2);
            \draw[post] (em 2) -- (result 3); 
        \end{tikzpicture} 
    \end{center} 
    \caption{Indirect inference model calibration. The gray circle in the top row
    corresponds to the agent-based simulation step, while blue rectangles correspond to
    estimation steps. See main text for explanation.}
    \label{fig:calib} 
\end{figure}

Figure \ref{fig:calib} summarizes our method. We first simulate from the
computer model (gray circle, top row) using a design with $N$ points
$\boldsymbol\theta_1, \dots, \boldsymbol\theta_N$, $\forall
i\;\boldsymbol\theta_i\in \mathbb R^p$. The points are chosen using Latin
Hypercube Sampling (from now on \textsc{lhs}; see Sec. \ref{ss:lhs}). From the
simulation step we obtain synthetic lifespan samples $\boldsymbol\tau_1, \dots,
\boldsymbol\tau_N$, where each sample contains a variable number of observations
$M_i = \left|\boldsymbol\tau_i\right|$. We then fit a \textsc{gmm} (blue
rectangle, top row) to each sample $\boldsymbol\tau_i$ and obtain the
corresponding auxiliary parameters vector $\boldsymbol{\hat\beta_i}$. Taken
together, the $N$ auxiliary parameter vectors form the training set $\boldsymbol
B_\mathrm{known}$ for the \textsc{gp}. 

We then apply the \textsc{gp} approximation (blue circle, middle row). Given an
untested computer model parameter vector $\boldsymbol\theta$, the \textsc{gp}
gives $\boldsymbol{\hat\beta}_\mathrm{GP}\left( \boldsymbol\theta \right) =
E\left[ \boldsymbol\beta\left( \boldsymbol\theta \right) | \boldsymbol
B_\mathrm{known} \right]$. This is an approximation of the unknown mapping
$\boldsymbol\beta\left( \boldsymbol\theta \right)$ between the agent-based model
parameters $\boldsymbol\theta$ and the parameters of the \textsc{gmm}
$\boldsymbol\beta$.

Separately, we fit the empirical dataset $\boldsymbol\tau$ to the \textsc{gmm},
and obtain the estimated parameters $\boldsymbol{\hat\beta}$ (bottom row).
Indirect inference then compares this information with
$\boldsymbol{\hat\beta}_\mathrm{GP}\left( \boldsymbol\theta \right)$, to find the
value of the agent-based model parameter $\boldsymbol\theta$ that gives the best
description of the empirical data.

\subsection{Indirect inference for model calibration}\label{ss:indinf}

Let us consider a generative model $\mathcal M$ with $p$ unknown parameters
$\boldsymbol \theta = \left( \theta_1, \theta_2, \ldots, \theta_p \right)$, and
$n$ independent, identically distributed observations from an empirical process
$\boldsymbol x = \left( x_1, x_2, \cdots, x_n \right)$. The IID assumption is
required by indirect inference, and for the case of activity lifespan data from
different individuals, like in our case, is easily satisfied. We assume that
maximum likelihood estimation of $\mathcal M$ is either intractable or that the
likelihood function $\mathcal{L}$ is unavailable in analytic form -- a common
case for agent-based models. However, $\mathcal M$ is generative and thus we can
simulate from it. How can we estimate this model then? Indirect inference
proposes an ingenious way to solve this problem.

Let us consider an \emph{auxiliary} model $\mathcal M_a$ with parameters
$\boldsymbol\beta$. The auxiliary model must foremost be  easy to fit to the data
$\boldsymbol x$. Intuitively, if the auxiliary model $\mathcal M_a$ is able to
capture the main feature of the data, that is, if it is sensitive enough to
changes of $\boldsymbol\theta$, then it induces an invertible function
$\boldsymbol\beta\left( \boldsymbol\theta \right)$ of the parameters of our
model. Estimation then amounts just to inverting this function, so that we find
the value of $\boldsymbol\theta$ associated to the estimate
$\boldsymbol{\hat\beta}$. Under the assumption that the empirical data have been
generated by a `true' value $\boldsymbol\theta_0$, this is the estimate
$\boldsymbol{\hat\theta}$ of the parameter $\boldsymbol\theta_0$ under model
$\mathcal M$. 

There are different ways to do this. The one we use in this paper is the so-called
``Wald approach'' to indirect inference, which minimizes the following quadratic
form:

\begin{equation} \hat{\boldsymbol\theta}_\mathrm{ii} =
    \arg\min_{\boldsymbol\theta} \left( \boldsymbol{\hat\beta} -
    \boldsymbol{\hat\beta}\left( \boldsymbol\theta \right) \right)^{T}
    \boldsymbol W \left( \boldsymbol{\hat\beta} - \boldsymbol{\hat\beta}\left(
    \boldsymbol\theta \right) \right) \label{eq:wald} \end{equation}

\noindent where $\boldsymbol W$ is a positive definite matrix that is used to
give more or less weight to the auxiliary parameters \cite{Smith2008a}. If the
asymptotic distribution of $\boldsymbol{\hat\beta}$ is normal, a common trick to
enhance convergence is to generate via simulation $S$ different realizations of
the data $\boldsymbol x^{\left( \boldsymbol\theta \right)}_1, \ldots,
\boldsymbol x^{\left( \boldsymbol\theta \right)}_S$ for a given
$\boldsymbol\theta$, fit each of them to the auxiliary model, and then take the
sample average.

The choice of a good auxiliary model $\mathcal M_a$ is critical here. In the
calibration of our peer production model we performed several diagnostic checks
in order to ensure that the required condition on $\boldsymbol\beta\left(
\boldsymbol\theta \right)$ is satisfied. 

\subsection{Gaussian mixture models for dimensionality reduction}\label{ss:gmm}

In this paper, we use \textsc{gmm} for two purposes: first, we want a
general-purpose auxiliary model that is good at summarizing the salient
features of the lifetime distribution. Second, we want to reduce the
dimensionality of the model output so that we can apply the \textsc{gp}
emulator. 

Formally, the density of mixture model with $k$ components is given by a
weighted average of the densities of each component:

\begin{equation}
    p\left(x\right) = \sum_{j=1}^k \pi_j p_j\left(x;\boldsymbol\theta_j\right)
    \label{eq:mixturedens} 
\end{equation} 

where $\sum_j \pi_j = 1$. When $p_j \sim \mathcal N\left( \mu_j, \sigma_j^2
\right)$ for all $j$ we speak of a \emph{Gaussian} mixture model. The vector of
auxiliary parameters has dimensionality $3k-1$:

\begin{equation}
  \boldsymbol\beta = \left( \mu_1, \ldots, \mu_k, \sigma_1, \ldots, \sigma_k,
  \pi_1, \ldots, \pi_{k-1} \right).
  \label{eq:auxiliary}
\end{equation}

As we said, we measure lifespan as the time elapsed between the first and the
last edit of a user. This means that our data (both the empirical and the
synthetic ones, see Section \ref{ss:patterns}) are simultaneously right-censored
and left-truncated. The right censoring is a natural consequence of the finitude
of the observation window. The left-truncation is instead due to physical
constraints of the speed at which humans can interact with a computer
interface.\footnote{For example Malmgren et al. attempted an estimation of the
    minimum time it takes a human to send two emails consecutively
    \cite{Malmgren2008a}.}

For ease of analysis, before estimating the \textsc{gmm} on our data we can
transform them to a fully truncated sample; that is, both right- and
left-truncated. We can in fact assume that after $\tau_\mathrm{max}$ days of
inactivity a user will be permanently so and focus only on the inactive users.
Of course, we do the same also for the synthetic data produced by the simulator.
Following the literature, we took $\tau_\mathrm{max} = 6$ months
\cite{Wilkinson2008a}.\footnote{Results from a simple sensitivity test suggest
    that the choice of $\tau_\mathrm{max}$ does not impact on the result of the
fit if taken large enough, e.g. 1 month or more.}

For each Wikipedia, we tested both a regular \textsc{gmm} and a \emph{truncated}
one -- that is, a model that assigns zero probability to data outside the
observation window. The window was estimated from the minimum and maximum
observations. For the number of components $K$, we tried models with $k=2,3$. 

\subsection{Latin Hypercube Sampling}\label{ss:lhs}

The indirect inference technique requires us to perform simulations from the
model. To do so, we need to choose a \emph{design}, which is the sites of the
parameters space at which we want to evaluate or agent-based model. 

Following the literature on computer code emulation \cite{Morris1995,Dancik2010a} we use a maximin Latin Hypercube design, an efficient,
space-filling, block design. A maximin design $\boldsymbol\Theta = \left(
\boldsymbol\theta_1, \dots, \boldsymbol\theta_N \right)$ maximizes the minimum
distance between any pair of points; that is:

\begin{equation}
    \max_{\boldsymbol\Theta}{\min_{i<i'} \big\|\boldsymbol\theta_i - \boldsymbol \theta_{i'} \big\|.}
    \label{eq:maxmin}
\end{equation}

In practice, for each community we sampled $10^4$ designs with $N=32$ and chose
the one that maximized Eq. \eqref{eq:maxmin}. For the Italian, French, and Portuguese
wikis, simulations of each site of the hypercube were repeated 10 times. 

With the exception of the parameters on which we are going to perform the
indirect inference, all other input variables of the model must be set to some
value that allows the response of the model to be compared with the empirical
data in the best possible way -- we see how in the next section.

\subsection{Estimation of additional parameters}\label{s:other}

\begin{table}[tbp]
    \footnotesize
    \centering
    \caption{Parameters for the simulations of the agent-based model of peer
        production. The column entitled ``Values''  may indicate the sampling interval
        used in the calibration procedure, or its value.}
    \label{tab:gsaparams}
    \begin{tabular}{@{\extracolsep{\fill}}lccr@{\extracolsep{0pt}}}
        \toprule
        Parameter & Symbol & Values & Unit \\
        \midrule
        Popularity dampening const. & $c_p$ & $(0, 100)$ & \\
        Initial motivation          & $c_s$ & $(0, 100)$ & \\
        Confidence bound            & $\varepsilon$ & $(0, \nf{1}{2})$ & \\
        Rollback probability        & $p_{\mathrm{rollback}}$ & $(0, 1)$ & \\
        Speed                       & $\mu$ & $(0, \nf{1}{2})$ & \\
        Daily sessions rate         & $\lambda_a$ & 1 & $\nf{1}{day}$ \\
        Session editing rate        & $\lambda_e$ & 1 & \nf{1}{\minute} \\
        Additional session edits    & $N_a$ & 1 & \\
        Daily rate of new pages     & $\rho_p$ & see Tab. \ref{tab:fixedparams} & $\nf{1}{day}$ \\
        Daily rate of new users     & $\rho_u$ & \textquotedbl & $\nf{1}{day}$ \\
        Long-term time scale        & $\tau_0$ & \textquotedbl & day \\
        Short-term time scale       & $\tau_1$ & \textquotedbl & day \\
        Simulation time             & $T$ & \textquotedbl & year \\
        \bottomrule
    \end{tabular}
\end{table}

The model presented in Section \ref{s:model} contains several parameters. Table
\ref{tab:gsaparams} summarizes them. For the purpose of estimation, we can
identify three types of parameters: those related to the editing cascades model
($N_a, \lambda_a, \lambda_e$), parameters that can be directly estimated from
the raw data (see Table \ref{tab:data}), and parameters that need to be
estimated via calibration.

The first group of parameters is not going to affect the distribution of $\tau$
too much, provided that a non-pathological choice is taken (e.g. a pathological
choice would be $N_a = 0$, which would recover the original Poisson process). In
particular, we need that users do at least two edits per session; that is, $N_a
\ge 1$. Provided this is the case, by the definition of $\tau$ any additional
edit will not affect the final lifespan of the users. Thus we can set $N_a = 1$
to save \textsc{cpu} cycles. Similarly, we need users to make at least one edit
session per day (i.e. $\lambda_e = 1 ~\textrm{day}^{-1}$), and that the rate of
activity within a session is set to a plausible value, such as one edit per
minute.

The parameters of the second group govern the microscopic dynamic of the model;
for example, the parameters for the update rules Eq. \eqref{eq:mod:1} and Eq.
\eqref{eq:mod:2}, or the popularity dampening factor of the page selection model
($c_p$). These cannot be readily estimated from the data on user activity and
thus are calibrated with the indirect inference technique showed before. For
these parameters, Table \ref{tab:gsaparams} reports the intervals used in the
Latin Hypercube Sampling scheme.

The parameters from the third group can be estimated from the raw data (see
Section \ref{s:data}). In particular, the simulation time interval $T$ can be
set to the obvious choice $T = t_1-t_0$ where $t_0$ and $t_1$ are the earliest
and the latest recorded time stamp in the data, respectively. In the following
we cover the estimation of the remaining parameters ($\rho_u$, $\rho_p$,
$\tau_0$, and $\tau_1$). Table \ref{tab:fixedparams} reports the results of
their estimation. 

\begin{table}[tbp]
    \footnotesize
    \centering
    \caption{Estimated parameters and uncertainties (s.d.) from raw data (see
        sec. \ref{s:data}) used in the calibration simulations.} 
    \label{tab:fixedparams}
    \begin{tabular}{@{\extracolsep{\fill}}ccr @{ $\pm$
        }lccc@{\extracolsep{0pt}}} \toprule Language & $\rho_u$ &
        \multicolumn{2}{c}{$\rho_p$} & $\tau_0$ & $\tau_1$ & $T$ \\
        & (\reciprocal\dday) & \multicolumn{2}{c}{(\reciprocal\dday)} &
        (\minute) & (\dday) & (\dday) \\
        \midrule
        Portuguese & $9.40\pm 0.14$  & $6.55$ & $0.61$ & $17.12 \pm 23.61$ &
        $1.01\pm 8.70\times 10^3$ & $3.03 \times 10^3$ \\ 
        Italian & $4.06\pm 0.07$ & $6.4$ & $0.13 \times 10^2$ & $15.43 \pm
        23.67$ & $1.35\pm 7.67\times 10^3$ & $2.96\times 10^3$ \\
        French & $2.19\pm 0.04$ & $1.11$ & $0.03\times 10^3$ & $15.81\pm 25.67$
        & $1.32\pm 5.72\times 10^3$ & $3.04\times 10^3$ \\ 
        \bottomrule 
    \end{tabular} 
\end{table}

\subsubsection{Time scales of user lifespan}\label{ss:timescales}

Perhaps the most important parameters we fit separately are the two time scales
$\tau_0$ and $\tau_1$. These are used to compute the activity lifespan
of any user during the simulations and, as we saw from
the factor screening, have an important impact on the overall lifespan
statistics. 

The approach we took was to estimate these two values directly from the data that we
wished to fit via indirect inference. Ideally, given a clustering of the user in
two classes, the short-term users and the long-term users, both $\tau_0$ and
$\tau_1$ should be computable from the observed user lifespans $\tau$ and the
information of group membership given by the latent variable computed by
\textsc{em} -- the so-called ``responsibilities''. In practice, instead of
running \textsc{em} we can take a much simpler approach and just define a hard
threshold $\tau_t$ for the lifespan of a user: observations of $\tau$ that are
less than $\tau_t$ are assigned to the cluster of short-lived users, while
observations of $\tau>\tau_t$ are assigned to the long-lived one. 

\begin{figure}[t]
  \begin{center}
    \includegraphics[width=.6\textwidth]{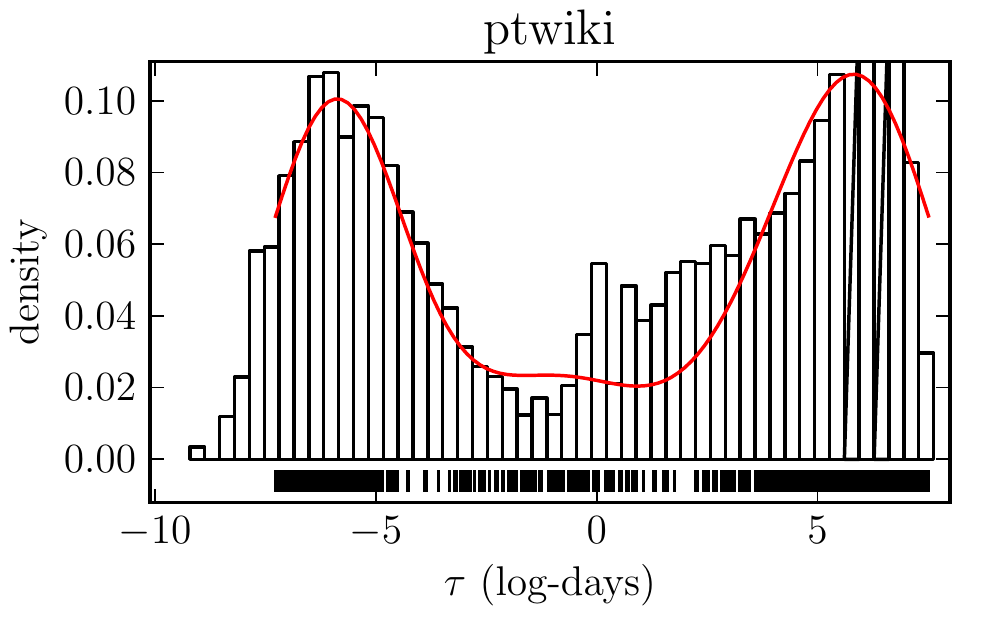}
  \end{center}
  \caption{Test simulation for the estimation of the user lifespan scales
  $\tau_{0,1}$. Histograms: empirical data from the Portuguese Wikipedia.
  Black bottom vertical lines: simulated observations. Red line: nonparametric
  (i.e. kernel) density estimate of simulated data. Simulation was performed
  with mean of clustered data and $\tau_t = 3\mathrm{h}$. Confidence
  $\varepsilon = 0.24$. Other parameters estimated from data of the Portuguese
  Wikipedia.}
  \label{fig:pttest}
\end{figure}

Once we have performed this (rather crude) form of hard clustering, we compute
suitable descriptive statistics that we take as the estimates for our two
parameters, that is, $\hat\tau_0$ and $\hat\tau_1$. We tested several values of
$\tau_t$ and eventually settled for $\tau_t = 3 \mathrm{h}$. As for the
statistics we used, we computed both median and mean activity lifespan of each
log-normal component. Our objective was to match the user activity lifespans
produced by our model, and therefore we performed some simulations with both values,
adjusting the value of $\varepsilon$ by hand, and found that the mean provided a
more reasonable estimate (see Fig. \ref{fig:pttest}). Larger values of the
threshold produce poor-quality estimates both for the mean and the median.

\subsubsection{Activity rates}\label{ss:activityrates}

Here we want to quantify the rates of activity for two processes: the arrival of
new users and the creation of new pages. In our model these two processes are
Poisson processes with a homogeneous rate of activity. Thus, our objective here
is to quantify the average activity rate for both processes, so that we set the
right scale of both processes for our calibration simulations. We estimate both
rates from data. 

To compute the rate of new users joining the community, we would need the daily
rate of new account creation. Unfortunately, the time stamps for the creation of the
accounts are not released to the public for obvious privacy reasons. As an
approximation, we use the time stamp of the first edit. The daily volume of new
users is then simply obtained by binning by date. Once we have the daily
volumes, we just compute $\hat\rho_u = \overline\rho$. A similar method was used
for estimating $\rho_p$ using the time stamp at which the page was created. This
is simply the time stamp of the first contribution, which can be obtained from
our raw data (see Section \ref{s:data}) by grouping by page \textsc{id} instead
of by user \textsc{id}.

\section{Results}
\label{s:results}

For each Wikipedia in our dataset we sampled a maximin \textsc{lhs} design with
50 points, estimated the auxiliary parameters of the \textsc{gmm}, and estimated
the parameters of the \textsc{GP} emulator. For the choice of the auxiliary
model, we tested both mixtures of $k=2,3$ components, with and without data
truncation. Before performing the final step in Eq. \eqref{eq:wald}, we computed
several diagnostic measures for each step of the calibration technique outlined
in Figure \ref{fig:calib}. 

\subsection{Approximation of auxiliary parameters via Gaussian process}

The choice of a good auxiliary model is important because
$\boldsymbol\beta\left( \boldsymbol\theta \right)$ must be able to capture the
features of the data well enough to be able to discriminate between different
choices of the parameters of the agent-based model $\boldsymbol\theta$. 

Identifiability may be hindered if multiple values of $\boldsymbol\theta$ result
in similar values of $\boldsymbol\beta$. A quick way to check this is to plot
$\boldsymbol\beta$, parametrized by $\boldsymbol\theta$, and see if the curve
crosses over itself at one or more points, that is, if $\exists
\boldsymbol\theta',\boldsymbol\theta'', \boldsymbol\theta' \neq
\boldsymbol\theta''$ s.t. $\boldsymbol\beta\left( \boldsymbol\theta' \right) =
\boldsymbol\beta\left( \boldsymbol\theta'' \right)$.

\begin{figure}[tbp]
  \includegraphics[width=\textwidth]{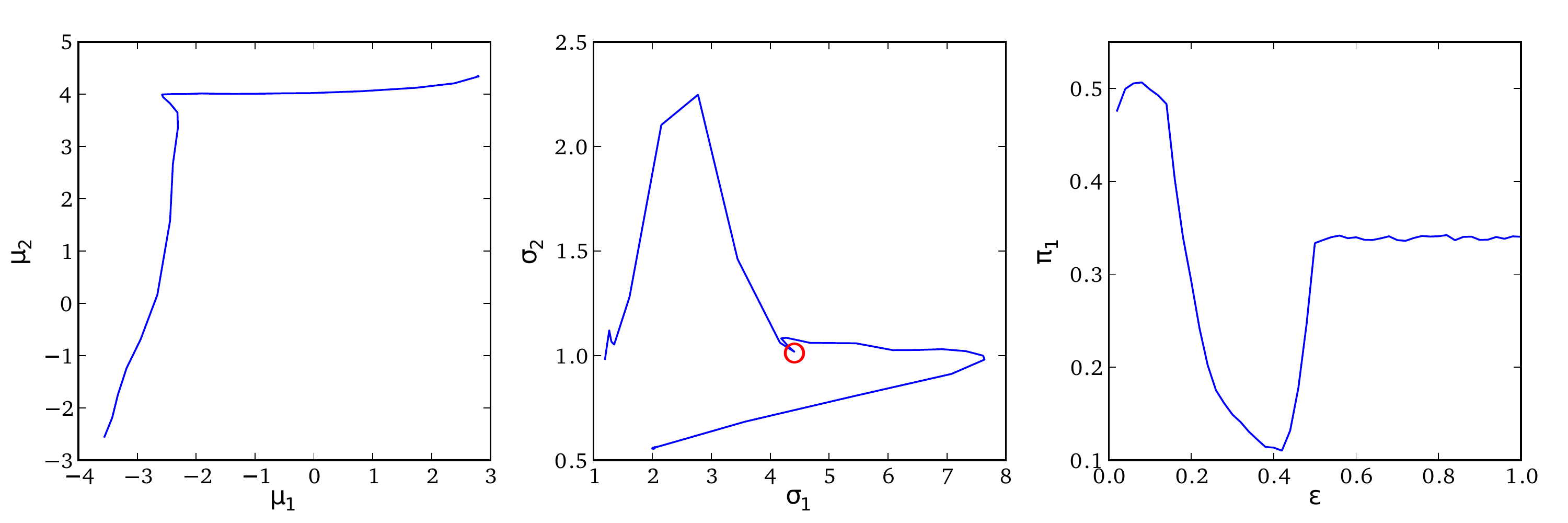}
  \caption{Auxiliary parameters as a function of $\varepsilon$. Auxiliary
  model: \textsc{gmm} with 2 components. Implicit and explicit parametric
  plots of parameter $\varepsilon$. Left: $\mu_1$ versus $\mu_2$; center:
  $\sigma_1$ versus $\sigma_2$; and right: $\pi_1$ versus $\varepsilon$.}
  \label{fig:paramplot}
\end{figure}

For illustrative purpose, we assessed the identifiability of the \textsc{gmm}
model on our model of peer production. Figure \ref{fig:paramplot} reports the
result of the test. We let $\varepsilon$ range in the interval $\left( 0,1
\right)$ and used the original (i.e. without edit cascades) peer production
model to produce parametric plots of the parameters of the auxiliary model
$\boldsymbol\beta\left( \varepsilon \right)$, as a function of $\varepsilon$.
The auxiliary model used in Fig. \ref{fig:paramplot} was a simple \textsc{gmm}
with two components, for a total of five parameters: two means ($\mu_1$ and
$\mu_2$), two variances ($\sigma_1$ and $\sigma_2$), and one weighting
coefficient ($\pi_1$). We produced graphs only for the three most meaningful
combinations of the ten possible pairwise choices of these parameters. The first
two plots ($\mu_1$ vs. $\mu_2$ and $\sigma_1$ vs. $\sigma_2$) are parameterized
implicitly by $\varepsilon$, while the last one reports the behavior of $\pi_1$
versus $\varepsilon$ explicitly. Close inspection of Figure \ref{fig:paramplot}
does not show any evident cross-overs. Cusps like the one present in the center
plot (circled red), are less critical for identifiability, but may still cause
computational problems if a local optimization technique to estimate
the auxiliary parameters is used. Other cusps may be present in the plots for
the other combinations, which we did not produce for this simpler model; our
estimation technique for the auxiliary model, however, is based on Expectation
Maximization and thus the presence of cusps should not present significant
problems.

\begin{figure}[tbp]
  \includegraphics[width=\textwidth,bb=100 100 864 864]{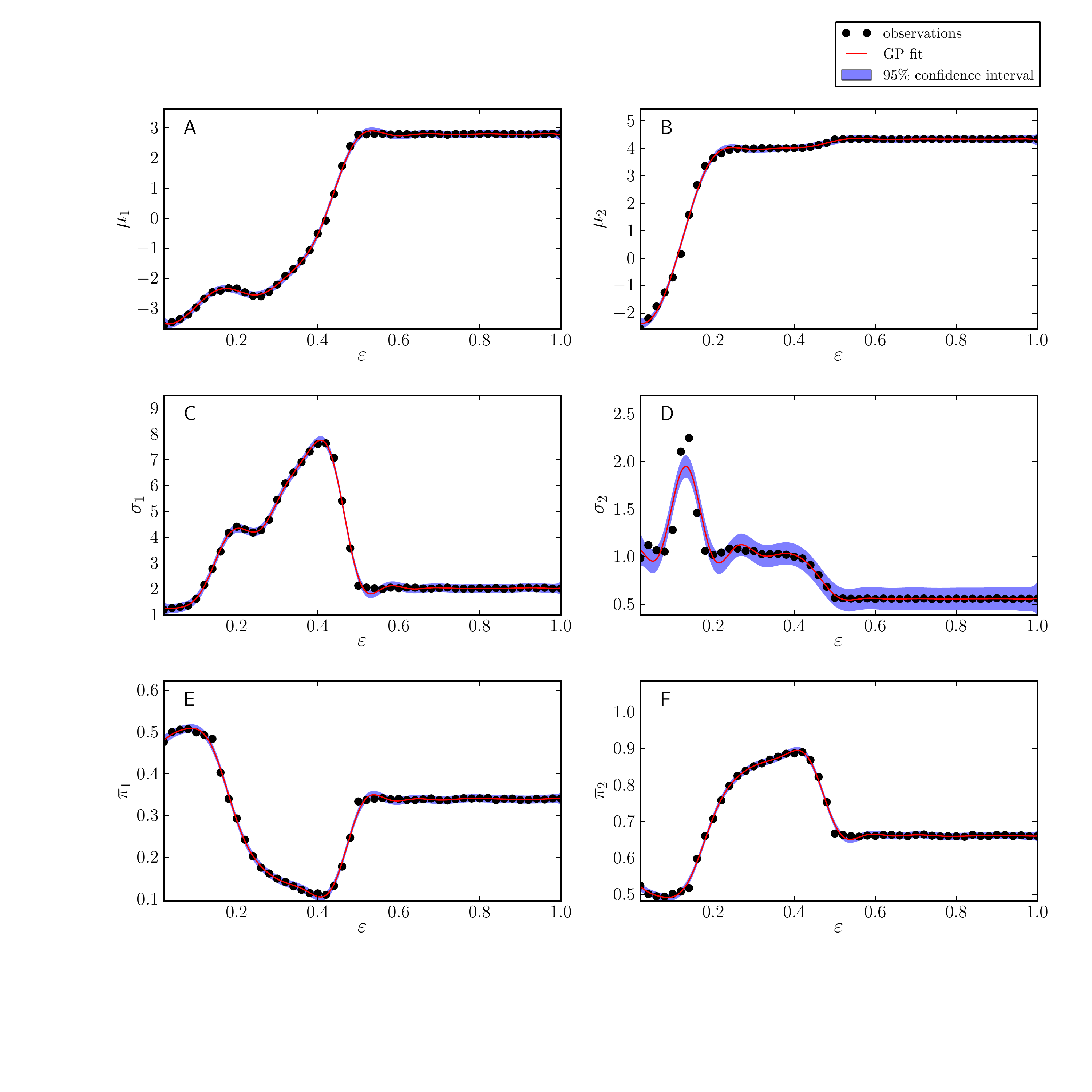}
  \caption{Gaussian Process approximation of auxiliary model. The
      \textsc{abm} model was simulated at 50 evenly spaced values of
      $\varepsilon\in\left[ 0,1 \right]$, \textsc{gmm} parameters ($k=2$) were
      estimated and fit to a univariate \textsc{gp}. $x$-axis: bounded
      confidence $\varepsilon$, $y$-axis: \textsc{gmm} parameter. Red line:
      \textsc{gp} fit. Black dots: estimated parameter. Purple area: 95\%
  confidence interval.} 
  \label{fig:gpfit} 
\end{figure}
 
The next step is to assess the \textsc{gp} approximation. We fit one Gaussian
process for each parameter of the \textsc{gmm}. In Figure \ref{fig:gpfit} we
show the result the \textsc{gp} approximation based on a learning set of 50
vectors of \textsc{gmm} parameters obtained from as many runs of the agent-based
model. In the figure, the blue area represents the 95\% confidence interval of
the approximation, which shows a very tight approximation. Note that there is no
change in the lifespan distribution for $\varepsilon\ge 0.5$, which is due to
the bounded confidence dynamics: in $\left[ 0,1 \right]$ the dynamics of
agreement always result in a full consensus case when $\varepsilon\ge
\nf{1}{2}$, and thus the average lifespan of the population is $\tau_0$
regardless of the value of $\varepsilon$ \cite{Fortunato2004}.

\subsection{Sensitivity analysis of auxiliary parameters}\label{s:sa}

\begin{table}
  \footnotesize
  \centering
  \caption{Decomposition of variance, \textsc{gmm}, $k=3$. Portuguese Wikipedia.
  Top panel: main effects. Lower panel: total interaction effects.}
  \label{tab:gmmvardec:pt}
  \begin{tabular}{@{\extracolsep{\fill}}crccccc@{\extracolsep{0pt}}}
    \toprule
    Variable & Variance & $p_{\mathrm{rollback}}$ & $\mu$ & $\varepsilon$ & $c_s$ & $c_p$\\
    \midrule
    $\mu_1$ & 1.503 & 0.023 & 0.038 & 0.389 & 0.095 & 0.020\\
    $\mu_2$ & 16.618 & 0.036 & 0.012 & 0.709 & 0.022 & 0.018\\
    $\mu_3$ & 11.331 & 0.031 & 0.028 & 0.698 & 0.055 & 0.027\\
    $\sigma_1$ & 0.970 & 0.014 & 0.029 & 0.369 & 0.153 & 0.026\\
    $\sigma_2$ & 1.288 & 0.016 & 0.044 & 0.309 & 0.125 & 0.068\\
    $\sigma_3$ & 0.071 & 0.041 & 0.025 & 0.499 & 0.055 & 0.018\\
    $\pi_1$ & 0.015 & 0.118 & 0.012 & 0.145 & 0.050 & 0.089\\
    $\pi_2$ & 0.039 & 0.054 & 0.014 & 0.304 & 0.232 & 0.038\\
    \midrule
    $\mu_1$ & 1.503 & 0.148 & 0.227 & 0.608 & 0.378 & 0.096\\
    $\mu_2$ & 16.618 & 0.115 & 0.084 & 0.845 & 0.138 & 0.078\\
    $\mu_3$ & 11.331 & 0.124 & 0.150 & 0.780 & 0.116 & 0.086\\
    $\sigma_1$ & 0.970 & 0.151 & 0.186 & 0.585 & 0.399 & 0.109\\
    $\sigma_2$ & 1.288 & 0.264 & 0.337 & 0.465 & 0.272 & 0.306\\
    $\sigma_3$ & 0.071 & 0.168 & 0.157 & 0.733 & 0.241 & 0.122\\
    $\pi_1$ & 0.015 & 0.290 & 0.259 & 0.506 & 0.288 & 0.382\\
    $\pi_2$ & 0.039 & 0.202 & 0.228 & 0.492 & 0.363 & 0.236\\
    \bottomrule
  \end{tabular}
\end{table}

\begin{table}
  \footnotesize
  \centering
  \caption{Decomposition of variance, truncated \textsc{gmm}, $k=2$. Italian
  Wikipedia. Top panel: main effects. Lower panel: total interaction effects.}
  \label{tab:tgmmvardec:it}
  \begin{tabular}{@{\extracolsep{\fill}}crccccc@{\extracolsep{0pt}}}
    \toprule
    Variable & Variance & $p_{\mathrm{rollback}}$ & $\mu$ & $\varepsilon$ & $c_s$ & $c_p$\\
    \midrule
    $\mu_1$ & 1.312 & 0.075 & 0.073 & 0.270 & 0.133 & 0.060\\
    $\mu_2$ & 2.845 & 0.037 & 0.032 & 0.771 & 0.043 & 0.014\\
    $\sigma_1$ & 0.533 & 0.077 & 0.076 & 0.281 & 0.139 & 0.057\\
    $\sigma_2$ & 0.018 & 0.019 & 0.088 & 0.138 & 0.097 & 0.073\\
    $\pi_1$ & 0.042 & -0.006 & 0.009 & 0.779 & 0.105 & 0.007\\[.5em]
    \midrule
    $\mu_1$ & 1.312 & 0.209 & 0.384 & 0.477 & 0.377 & 0.088\\
    $\mu_2$ & 2.845 & 0.107 & 0.048 & 0.866 & 0.108 & 0.054\\
    $\sigma_1$ & 0.533 & 0.214 & 0.355 & 0.479 & 0.375 & 0.085\\
    $\sigma_2$ & 0.018 & 0.188 & 0.423 & 0.435 & 0.489 & 0.182\\
    $\pi_1$ & 0.042 & 0.032 & 0.068 & 0.805 & 0.171 & 0.041\\
    \bottomrule
  \end{tabular}
\end{table}

\begin{table}
  \footnotesize
  \centering
  \caption{Decomposition of variance, \textsc{gmm}. French Wikipedia. Top panel:
  main effects. Lower panel: total interaction effects.}
  \label{tab:gmmvardec:fr}
  \begin{tabular}{@{\extracolsep{\fill}}crccccc@{\extracolsep{0pt}}}
    \toprule
    Variable & Variance & $p_{\mathrm{rollback}}$ & $\mu$ & $\varepsilon$ & $c_s$ & $c_p$\\
    \midrule
    $\mu_1$ & 0.511 & 0.012 & 0.037 & 0.294 & 0.127 & -0.014\\
    $\mu_2$ & 2.323 & -0.001 & -0.002 & 0.819 & 0.007 & -0.014\\
    $\sigma_1$ & 0.470 & 0.018 & 0.044 & 0.295 & 0.144 & -0.008\\
    $\sigma_2$ & 0.020 & 0.001 & 0.037 & 0.165 & 0.043 & 0.010\\
    $\pi_1$ & 0.041 & -0.009 & -0.013 & 0.793 & 0.090 & -0.002\\[.5em]
    \midrule
    $\mu_1$ & 0.511 & 0.152 & 0.317 & 0.534 & 0.449 & 0.081\\
    $\mu_2$ & 2.323 & 0.065 & 0.034 & 0.925 & 0.102 & 0.049\\
    $\sigma_1$ & 0.470 & 0.161 & 0.280 & 0.521 & 0.458 & 0.070\\
    $\sigma_2$ & 0.020 & 0.187 & 0.328 & 0.570 & 0.573 & 0.114\\
    $\pi_1$ & 0.041 & 0.030 & 0.057 & 0.846 & 0.166 & 0.034\\
    \bottomrule
  \end{tabular}
\end{table}

For a more quantitative assessment of the \textsc{gmm} model as an auxiliary
model, we saw how sensitive each auxiliary parameter was to changes in the
inputs of the agent-based model. This is essentially a factor screening
exercise. Moreover, sensitivity indices can also be used to define the matrix
$\boldsymbol W$ of eq. \eqref{eq:wald}. We performed a global sensitivity
analysis of the auxiliary parameters. We computed main and total interaction
effect indices using a decomposition of variance based on the winding stairs
method \cite{Chan2000a,Saltelli2004a}.  

We performed the sensitivity analyses for each type of \textsc{gmm} and for
each dataset. For convenience here we report only the results for the choice of
the auxiliary model that we effectively used later in the calibration (tables
\ref{tab:gmmvardec:pt}--\ref{tab:gmmvardec:fr}), but found nonetheless similar
results for the other combinations. Small negative indices near zero were due to
the sampling uncertainty in the winding stairs method. 

The result of the sensitivity analysis shows that the auxiliary parameters with
the highest variance were the locations $\mu$ of the mixture components (i.e.
the means) and, to some extent, the variances $\sigma$. A truncated model seemed
to decrease the difference in variability between the mixture means. Most
importantly, main effect indices show that the confidence parameter
($\varepsilon$) was  responsible for most of the variability of the auxiliary
parameters. In summary, the auxiliary model was especially sensitive to changes
in $\varepsilon$, weakly sensitive to changes in $c_s$, and almost insensitive
to other parameters.

\subsection{Cross-validation}\label{s:cv}

\begin{table}[tbp]
    \footnotesize
    \centering
    \caption{Results of leave-one-out cross-validation. Coefficient of
        determination $R^2$. For each language, the best $R^2$ attained over
        parameter $\varepsilon$ is shown in bold.}
    \label{tab:cvresults}
    \begin{tabular}{@{\extracolsep{\fill}}lccccc@{\extracolsep{0pt}}}
        \toprule
        & $\mu$ & $\varepsilon$ & $p_{\mathrm{rollback}}$ & $c_s$ & $c_p$ \\
        \midrule
        \multicolumn{6}{c}{Portuguese}\\
        \midrule
        {\sc gmm}, $k=2$ & 0.02 & 0.73 & 0.00 & 0.13 & 0.02 \\
        {\sc gmm}, $k=3$ & 0.03 & {\bf 0.86} & 0.16 & 0.02 & 0.01 \\
        {\sc tgmm}, $k=2$ & 0.01 & 0.70 & 0.01 & 0.36 & 0.01 \\
        {\sc tgmm}, $k=3$ & 0.04 & 0.85 & 0.00 & 0.28 & 0.01 \\
        \midrule
        \multicolumn{6}{c}{Italian}\\
        \midrule
        {\sc gmm}, $k=2$ & 0.00 & 0.91 & 0.01 & 0.66 & 0.09 \\
        {\sc gmm}, $k=3$ & 0.02 & 0.90 & 0.01 & 0.30 & 0.03 \\
        {\sc tgmm}, $k=2$ & 0.02 & {\bf 0.93} & 0.03 & 0.75 & 0.01 \\
        {\sc tgmm}, $k=3$ & 0.00 & 0.85 & 0.00 & 0.42 & 0.03 \\
        \midrule
        \multicolumn{6}{c}{French}\\
        \midrule
        {\sc gmm}, $k=2$ & 0.00 & {\bf 0.91} & 0.01 & 0.61 & 0.04 \\
        {\sc gmm}, $k=3$ & 0.01 & 0.90 & 0.03 & 0.33 & 0.01 \\
        {\sc tgmm}, $k=2$ & 0.01 & 0.76 & 0.00 & 0.69 & 0.08 \\
        {\sc tgmm}, $k=3$ & 0.08 & 0.86 & 0.16 & 0.35 & 0.09 \\
        \bottomrule
    \end{tabular}
\end{table}

\begin{figure}[tbp]
  \begin{center}
    \includegraphics[width=\textwidth,bb=50 80 533.973572 576]{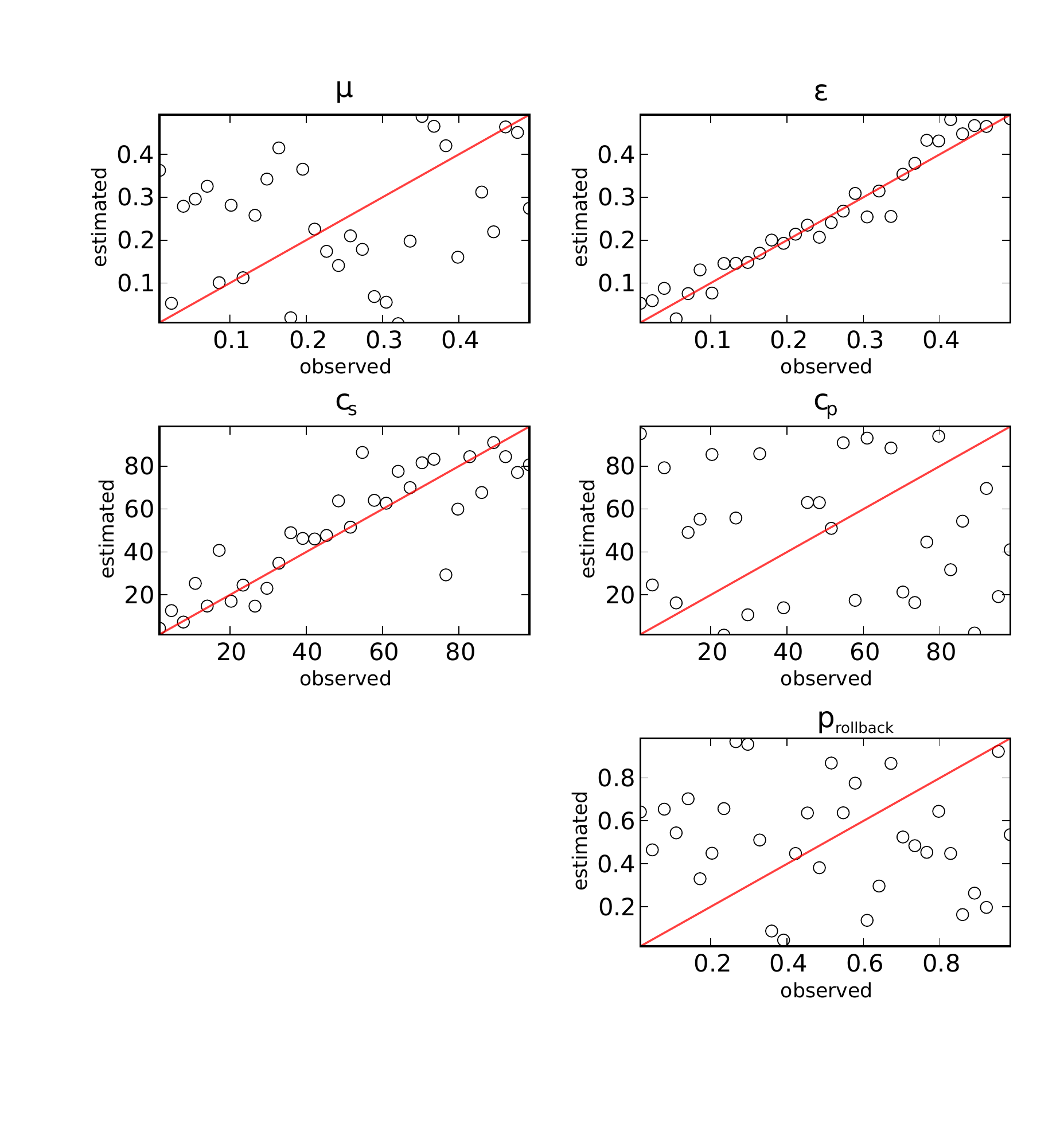}
  \end{center}
  \caption{Leave-one-out cross validation. $x$-axis: observed holdout parameter
      values obtained by \textsc{lhs}. $y$-axis: indirect inference estimates
      based on remaining data. Simulations were carried out with the
      parameterization for the Italian Wikipedia (see Table
      \ref{tab:fixedparams}); the auxiliary model was a truncated \textsc{gmm},
  $k=2$. The red line has slope equal to 1 and serves a guide to the eye.}
  \label{fig:cvtgmm2:it} \end{figure}

Finally, we used leave-one-out cross-validation to quantify the accuracy of
indirect inference in reconstructing the parameters of the model. Using
\textsc{lhs} we sampled $N$ vectors of parameters and simulated from the
agent-based model to obtain as many samples of user activity lifespans. We then
set aside one pair $\left( \boldsymbol\theta_h, \boldsymbol\tau_h \right)$ as
test set (the holdout) and used the remaining $N-1$ pairs as a learning set for
the indirect inference technique, which we used to estimate
$\boldsymbol{\hat\theta_h}$ from the synthetic lifespan data
$\boldsymbol\tau_h$. Repeating this exercise for $h=1\ldots N$, we can the plot
the estimated \textsc{abm} parameters as a function of the true ones, and
compute coefficient of determination $R^2$ to assess the performance of the
calibration. 

We performed several cross-validations, for each language and auxiliary model
combination. For each of those, we tested both a weighted and an unweighted
indirect inference technique. In the weighted case, the diagonal of the matrix
$\boldsymbol W$ from Eq. \eqref{eq:wald} was set to the variances of the
parameters from the global sensitivity analysis (see Tables
\ref{tab:gmmvardec:pt}--\ref{tab:gmmvardec:fr}), while in the unweighted case
$\boldsymbol W = \boldsymbol I$. Surprisingly, the best results were those with
no weighting, which are those we choose to report here. Table
\ref{tab:cvresults} reports the results.

While accuracy, as determined by the $R^2$, varied sometimes appreciably across
flavors of the auxiliary model and languages, it seemed consistent with the
results of the global sensitivity analysis. The best-performing auxiliary model
for each language was selected looking at the $R^2$ for the confidence parameter
(in bold in the table). As a graphical companion to the table, Figure
\ref{fig:cvtgmm2:it} shows the cross-validation plot for the Italian case. As
expected, the accuracy is excellent for the confidence parameter $\varepsilon$
and good for the initial motivation $c_s$, while for other parameters no clear
linear trend can be discerned.

\subsection{Indirect inference}

\begin{table}[tbp]
    \footnotesize
    \centering
    \caption{Calibrated parameters with standard errors.}
    \label{tab:fitres}
    \begin{tabular}{@{\extracolsep{\fill}}cccccc@{\extracolsep{0pt}}}
        \toprule
        wiki & $p_{\mathrm{rollback}}$ & $\mu$ & $\varepsilon$ & $c_s$ & $c_p$ \\
        \midrule
        pt & 0.52 \textpm 0.01 & 0.46 \textpm 0.00 & 0.39 \textpm 0.00 & 70.78 \textpm 0.80 & 51.56 \textpm 0.79 \\
        it & 0.36 \textpm 0.00 & 0.21 \textpm 0.00 & 0.49 \textpm 0.00 & 53.81 \textpm 0.61 & 58.31 \textpm 0.57 \\
        fr & 0.02 \textpm 0.01 & 0.02 \textpm 0.00 & 0.49 \textpm 0.00 & 3.79 \textpm 0.86 & 89.37 \textpm 0.77 \\
        \bottomrule
    \end{tabular}
\end{table}

\begin{figure}[tbp]
  \includegraphics[width=0.49\textwidth]{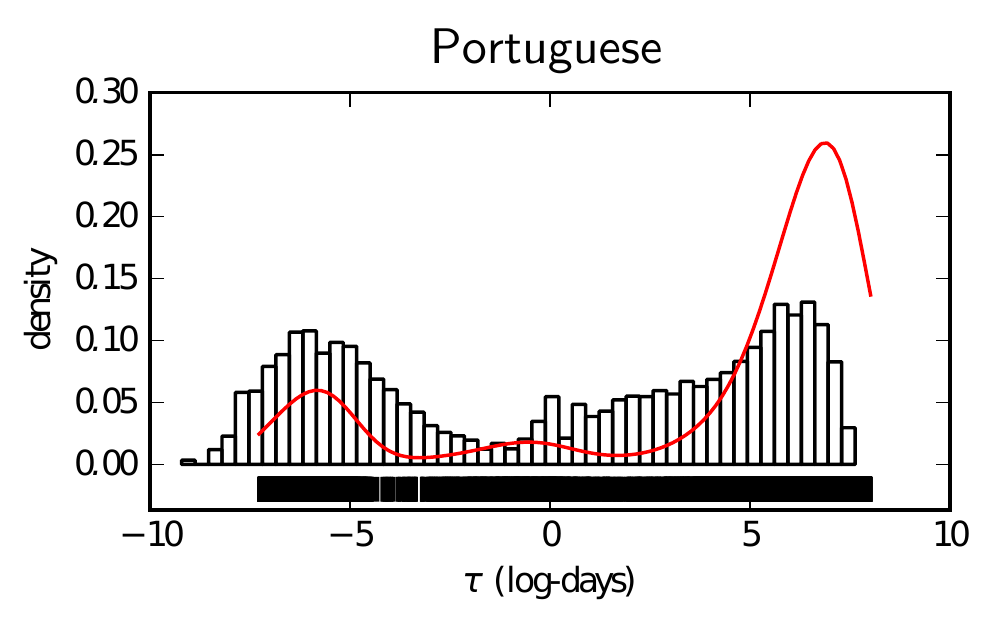}
  \includegraphics[width=0.49\textwidth]{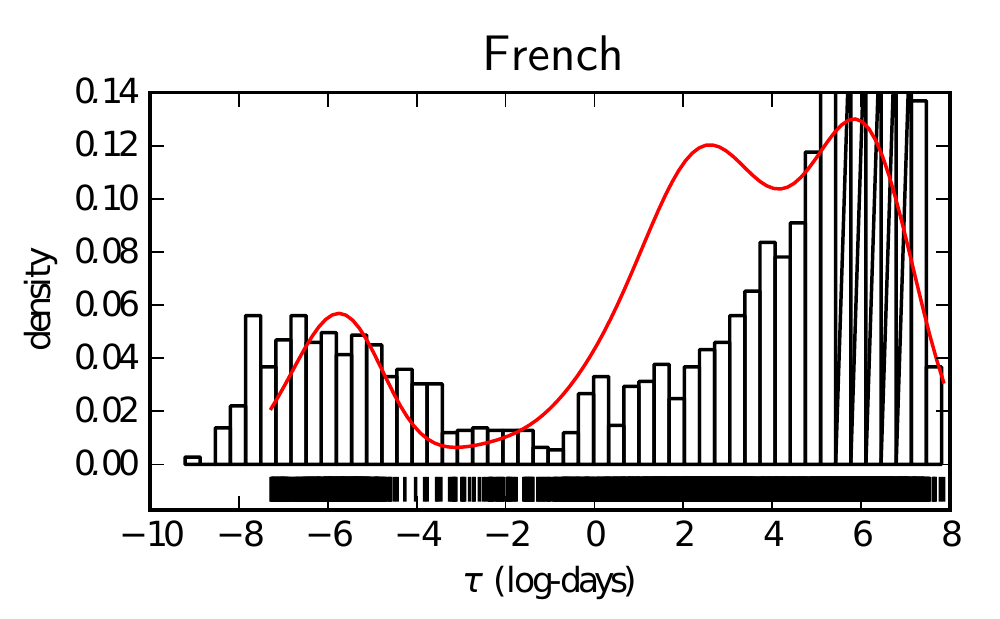}\\[1em]
      \includegraphics[width=0.49\textwidth]{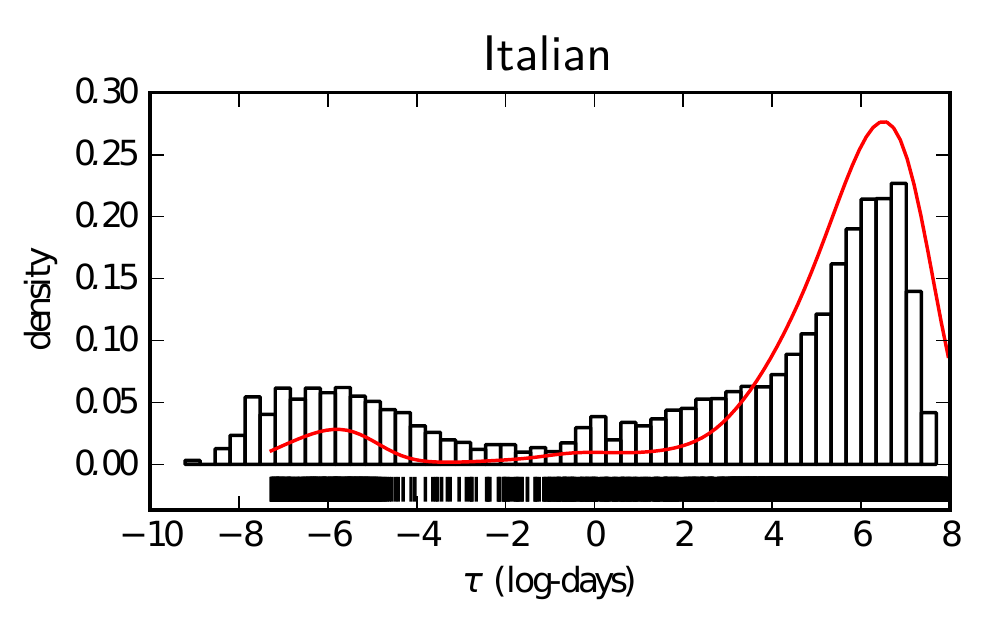}
  \caption{Comparison between empirical data (histograms) and simulated data
  (kernel density estimate, red line), obtained from the calibrated model.}
  \label{fig:fitresults}
\end{figure}

Having tested the accuracy of the indirect inference technique and selected the
best auxiliary models for each language, we finally applied it to the empirical
data to get estimates of the parameters of our model. Table \ref{tab:fitres}
reports the results of the calibration; standard errors were computed on a
bootstrapped sample with 1000 observations. 

With these, we simulated from the calibrated model. As a way to check the model
fit visually, we plotted in Figure \ref{fig:fitresults} a kernel density estimate of the synthetic data
together with histograms of empirical data. 

\section{Discussion}
\label{s:discussion}

Indirect inference is a promising framework for calibrating computational
models, but its application requires some care and is not as straightforward as
classic inference techniques such as maximum likelihood. In particular, it is
important to assess independently the performance of the auxiliary model in
capturing the salient aspects of the model. This is especially useful for
assessing the accuracy of the model in estimating the various model parameters.

Useful diagnostic tests include parametric plots for assessing potential
problems with identification (Fig. \ref{fig:paramplot}) and diagnostic plots of
the \textsc{gp} emulator (Fig. \ref{fig:gpfit}). Moreover, the results of the
cross-validation (Section \ref{s:cv}) were consistent with those of the global
sensitivity analysis of the auxiliary model (Section \ref{s:sa}), which shows
exactly how multiple independent methods can provide a comprehensive diagnostic
picture of a calibration method before one sets out to apply it to a model. 

Using the global sensitivity analysis and cross-validation we learned that the
\textsc{gmm} model could capture well two parameters out of the five we tried to
calibrate, namely the confidence bound $\varepsilon$ and the initial motivation
$c_s$ (see Fig \ref{fig:cvtgmm2:it}). While the result for $\varepsilon$ is
expected, that for $c_s$ comes a bit as an interesting surprise, and can be
explained by noting that this parameter is able to influence the location of the
short-term component of lifespan distribution, a feature that the mixture model
is indeed able to detect, albeit partially. 

The method is not reliable at reconstructing the other three parameters, namely
the speed $\mu$, the rollback probability $p_{\mathrm{rollback}}$, and the
popularity dampening factor $c_p$, as evidenced by the coefficient of
determination of the leave-one-out cross-validation. This means that we gained
little by calibrating these parameters, and could have set them to the mid-range
of their intervals before running the calibration simulations, thus calibrating
a simpler model. The low sensitivity of the auxiliary model to these parameters
may be due to the inability of the auxiliary model in capturing the patterns in
the data, or to the norm formation model itself; that is, the model may be
underdetermined by the data. 

Because the Gaussian mixture model is generally regarded as a versatile model
for complicate data, such as the user activity lifespan \cite{Ciampaglia2010a},
we deem the latter a more likely explanation than the former.

Taken together, these results highlight how the choice of a good auxiliary model
is critical to the successful application of indirect inference, and how factor
screening via global sensitivity analysis can be leveraged to assess the quality
of an auxiliary model. Conversely, one could also argue that the calibration
exercise, as outlined here, is useful for assessing whether, and especially
where, a model is underdetermined by the data, and thus it could point to an
effective methodology for identifying the parts that may be simplified. Thus
indirect inference could become a valuable tool in the toolbox of every
agent-based modeling practitioner.

If, however, one would be interested in estimating their value, then the
question remains open as to which approach should be used for calibrating
reliably the remaining three parameters mentioned above. A different auxiliary
model could be used, but the possibility of incorporating data variables other
than the activity lifespan should be taken into account, since it would
alleviate the under-determination problem. This is most pressing for a parameter
like $p_{\mathrm{rollback}}$, which is supposed to play an important role in the
mechanism of norm emergence, while it seems less critical for parameters like
$\mu$ and $c_p$; in the first case, this parameter is considered less important
for the dynamics of the \textsc{bc} model \cite{Castellano2009a}, while in the
other the corresponding mechanism (page selection) may be just too complicated
and thus could be dropped in favor of a simpler model.

Comparison of the prediction of the calibrated model with the empirical data
(Fig. \ref{fig:fitresults}) shows discrepancies at a range of very short time
scales ($\approx 10^{-8}\;\log\left( \dday \right)$), and a degree of
overestimation of $\varepsilon$, resulting in a distribution that gives too much
probability at large lifespan scales. Moreover, the estimates for the confidence
bound $\varepsilon$ parameter (Tab. \ref{tab:fitres}) are very close to the full
consensus value $\varepsilon = 0.5$; this is the value past which consensus is
always the case, suggesting an over-estimation of this parameter. This would be
in fact counter-intuitive, considered the general opinion that Wikipedia is a
conflict-laden arena, and one straightforward implication would also be the
absence of minority groups in the community of Wikipedia users, which seems also
in contrast with general observation. Table \ref{tab:fitres} also shows large
differences among the estimates for $c_s$. The calibration accuracy for this
parameter is lower than for $\varepsilon$, which might explain these big
differences. 

For such a simplistic model, the results are nonetheless encouraging. Better
fits could be attained with more sophisticate estimation of parameters such as
the time scales $\tau_0$ and $\tau_1$, as well as by introducing a more
sophisticated model of the distribution of the time between consecutive edits of
users \cite{Radicchi2009a,Wu2010}. The new parameters, if any, introduced by
such models could be estimated separately, or possibly included in the indirect
inference.

Evaluation of the model inadequacy, of the uncertainty in the predictions due to
use of a surrogate model and parameter estimation and goodness of fit measures
are of course all desirable improvements on the methods presented here
\cite{Kennedy2001,Higdon2005}. Statistical tests for goodness of fit exist for
the classic indirect inference technique \cite{Gourieroux1993a}; thus it would
be interesting to see how to apply them when an additional layer of emulation is
required, as it is in our case with the Gaussian Process.

\section{Conclusions}
\label{s:conclusions}

We have presented a method for calibrating an agent-based model of
the activity lifespan of users in a community of peer production. The method is
especially suited for models with high-dimensional outputs and long simulation
times, as  is often the case in social simulation.

Peer production communities provide a unique opportunity to study the emergence
of social norms. Norms, including social production norms, contribute to the
distinctive culture of an online community and thus the process of community
formation can be regarded more broadly as a process of cultural dynamics.
Several agent-based models have been proposed to explain various aspects of the
process of cultural formation in a generic setting
\cite{Axelrod1997a,Parisi2003,Klemm2003,Centola2007,Flache2011a}. It would be
interesting to know whether these models could be empirically validated in some
way, and calibration techniques such as the one presented here could prove
useful to this end.

\section*{Acknowledgments}

The author wishes to thank Alberto Vancheri and Paolo Giordano for support and
insightful discussions, Amirhossein Malekpour and Fabio Pedone for granting
access to their computing cluster, Erik Zachte for help with statistics about
Wikipedia, John McCurley for proofreading the manuscript, and the anonymous
reviewers, whose comments helped to improve the manuscript. The author
acknowledges the Swiss National Science Foundation for financial support by
means of project nr. 125128 ``Mathematical modeling of online communities''.

\bibliographystyle{plain}
\bibliography{abmcalibration}

\begin{thebibliography}{10}

\bibitem{Alexander2007}
J.~McKenzie Alexander.
\newblock {\em The Structural Evolution of Morality}.
\newblock Cambridge University Press, 2007.

\bibitem{Axelrod1986}
Robert Axelrod.
\newblock An evolutionary approach to norms.
\newblock {\em American Political Science Review}, 80:1095--1111, 1986.

\bibitem{Axelrod1997a}
Robert Axelrod.
\newblock {\em The complexity of cooperation}.
\newblock Princeton Studies in Complexity. Princeton University Press, 1997.

\bibitem{Barabasi1999}
Albert-L\'aszl\'o Barab\'asi and R\'eka Albert.
\newblock Emergence of scaling in random networks.
\newblock {\em Science}, 286(5439):509--512, 1999.

\bibitem{Barabasi2005a}
Albert-Lázló Barabási.
\newblock The origin of bursts and heavy tails in human dynamics.
\newblock {\em Nature}, 435(7039):207--211, 2005.

\bibitem{Bayarri2007}
Maria~J Bayarri, James~O Berger, Rui Paulo, Jerry Sacks, John~A Cafeo, James
  Cavendish, Chin-Hsu Lin, and Jian Tu.
\newblock A framework for validation of computer models.
\newblock {\em Technometrics}, 49(2):138--154, 2007.

\bibitem{Benkler2006}
Y.~Benkler.
\newblock {\em The wealth of networks: How social production transforms markets
  and freedom}.
\newblock Yale University Press, 2006.

\bibitem{Bianchi2007a}
Carlo Bianchi, Pasquale Cirillo, Mauro Gallegati, and Pietro Vagliasindi.
\newblock Validating and calibrating agent-based models: A case study.
\newblock {\em Computational Economics}, 30(3):245--264, 10 2007.

\bibitem{Bicchieri2005}
C.~Bicchieri.
\newblock {\em The grammar of society: The nature and dynamics of social
  norms}.
\newblock Cambridge University Press, 2005.

\bibitem{Bonabeau2002a}
Eric Bonabeau.
\newblock {Agent-based modeling: Methods and techniques for simulating human
  systems}.
\newblock {\em Proceedings of the National Academy of Sciences of the United
  States of America}, 99(Suppl 3):7280--7287, 2002.

\bibitem{Camerer2004}
Colin~F. Camerer and Ernst Fehr.
\newblock {\em Measuring Social Norms and Preferences Using Experimental Games:
  A Guide for Social Scientists}, chapter~3, pages 55--95.
\newblock Oxford University Press, 2004.

\bibitem{Castellano2009a}
Claudio Castellano, Santo Fortunato, and Vittorio Loreto.
\newblock Statistical physics of social dynamics.
\newblock {\em Rev. Mod. Phys.}, 81(2):591--646, May 2009.

\bibitem{Centola2007}
Damon Centola, Juan~Carlos González-Avella, Víctor~M. Eguíluz, and Maxi
  San~Miguel.
\newblock Homophily, cultural drift, and the co-evolution of cultural groups.
\newblock {\em Journal of Conflict Resolution}, 51(6):905--929, 2007.

\bibitem{Chan2000a}
Karen Chan, Andrea Saltelli, and Stefano Tarantola.
\newblock Winding stairs: A sampling tool to compute sensitivity indices.
\newblock {\em Statistics and Computing}, 10:187--196, 2000.

\bibitem{Cheshire2008}
Coye Cheshire and Judd Antin.
\newblock The social psychological effects of feedback on the production of
  internet information pools.
\newblock {\em Journal of Computer-Mediated Communication}, 13(3):705--727,
  2008.

\bibitem{Ciampaglia2011}
Giovanni~Luca Ciampaglia.
\newblock A bounded confidence approach to understand user participation in
  peer production systems.
\newblock In {\em Third International Conference on Social Informatics
  (SocInfo'11)}, LNCS, Singapore, 6 - 8 October 2011. Springer Verlang.

\bibitem{Ciampaglia2010a}
Giovanni~Luca Ciampaglia and Alberto Vancheri.
\newblock Empirical analysis of user participation in online communities: the
  case of {W}ikipedia.
\newblock In {\em Proceedings of ICWSM}, 2010.

\bibitem{Ciffolilli2003}
Andrea Ciffolilli.
\newblock Phantom authority, self-selective recruitment and retention of
  members in virtual communities: The case of wikipedia.
\newblock {\em First Monday}, 8(12), Dec 2003.

\bibitem{Clark2008}
William A.~V. Clark and Mark Fossett.
\newblock Understanding the social context of the {S}chelling segregation
  model.
\newblock {\em Proceedings of the National Academy of Sciences},
  105(11):4109--4114, 2008.

\bibitem{Dancik2010a}
Garrett~M. Dancik, Douglas~E. Jones, and Karin~S. Dorman.
\newblock Parameter estimation and sensitivity analysis in an agent-based model
  of {L}eishmania major infection.
\newblock {\em Journal of Theoretical Biology}, 262(3):398 -- 412, 2010.

\bibitem{Deci1999}
E.L. Deci, R.~Koestner, and R.M. Ryan.
\newblock A meta-analytic review of experiments examining the effects of
  extrinsic rewards on intrinsic motivation.
\newblock {\em Psychological bulletin}, 125(6):627, 1999.

\bibitem{Deffuant2002a}
G~Deffuant, F~Amblard, G~Weisbuch, and T~Faure.
\newblock How can extremism prevail? a study based on the relative agreement
  interaction model.
\newblock {\em J. Art. Soc. Soc. Sim.}, 5(4):paper 1, 2002.

\bibitem{Deffuant2001a}
Guillame Deffuant, David Neau, Frederic Amblard, and Gérard Weisbuch.
\newblock Mixing beliefs among interacting agents.
\newblock {\em Adv. Comp. Sys.}, 3:87--98, 2001.

\bibitem{Ellner2006a}
Stephen~P. Ellner and John Guckenheimer.
\newblock {\em Dynamic Models in Biology}.
\newblock Princeton University Press, 2006.

\bibitem{Epstein1996a}
Joshua~M Epstein and Robert Axtell.
\newblock {\em Growing artifical societies: social sciences from the bottom
  up}.
\newblock The MIT Press, Cambridge, MA, USA, 1996.

\bibitem{Feldman1984}
Daniel~C. Feldman.
\newblock The development and enforcement of group norms.
\newblock {\em The Academy of Management Review}, 9(1):pp. 47--53, 1984.

\bibitem{Flache2011a}
Andreas Flache and Michael~W. Macy.
\newblock Local convergence and global diversity: From interpersonal to social
  influence.
\newblock {\em Journal of Conflict Resolution}, 55(6):970--995, 2011.

\bibitem{Fortunato2004}
S.~Fortunato.
\newblock Universality of the threshold for complete consensus for the opinion
  dynamics of {D}effuant et al.
\newblock {\em International Journal of Modern Physics C}, 15:1301--1307, 2004.

\bibitem{Fortunato2007a}
Santo Fortunato and Claudio Castellano.
\newblock Scaling and universality in proportional elections.
\newblock {\em Phys. Rev. Lett.}, 99(13):138701, Sep 2007.

\bibitem{Gillespie1977a}
Daniel Gillespie.
\newblock Exact stochastic simulation of coupled chemical reactions.
\newblock {\em Journal of Physical Chemistry}, 81(25):2340--2361, 1977.

\bibitem{Gilli2003a}
M.~Gilli and P.~Winker.
\newblock A global optimization heuristic for estimating agent based models.
\newblock {\em Computational Statistics \& Data Analysis}, 42(3):299--312,
  March 2003.

\bibitem{Gourieroux1993a}
C~Gouri\'eroux, A~Monfort, and E~Renault.
\newblock Indirect inference.
\newblock {\em Journal Of Applied Econometrics}, 8({Suppl. S}):85--118, Dec
  1993.
\newblock {Conference On Econometric Inference Using Simulation Techniques,
  Rotterdam, Netherlands, Jun 05-06, 1992}.

\bibitem{Graham2003}
Charles~R. Graham.
\newblock A model of norm development for computer-mediated teamwork.
\newblock {\em Small Group Research}, 34(3):322--352, 2003.

\bibitem{Guo2009a}
Lei Guo, Enhua Tan, Songqing Chen, Xiaodong Zhang, and Yihong~(Eric) Zhao.
\newblock Analyzing patterns of user content generation in online social
  networks.
\newblock In {\em {KDD '09: Proceedings of the 15th ACM SIGKDD international
  conference on Knowledge discovery and data mining}}, pages 369--378, New
  York, NY, USA, 2009. ACM.

\bibitem{Halfaker2011}
Aaron Halfaker, Aniket Kittur, and John Riedl.
\newblock Don't bite the newbies: how reverts affect the quantity and quality
  of wikipedia work.
\newblock In {\em Proceedings of the 7th International Symposium on Wikis and
  Open Collaboration}, WikiSym '11, pages 163--172, New York, NY, USA, 2011.
  ACM.

\bibitem{Hamilton1964}
W.D. Hamilton.
\newblock The genetical evolution of social behaviour. i.
\newblock {\em Journal of Theoretical Biology}, 7(1):1--16, 1964.

\bibitem{Hardin1982}
Russell Hardin.
\newblock {\em Collective action}.
\newblock Resources for the Future. Johns Hopkins University Press, 1982.

\bibitem{Hegselmann2002a}
Rainer Hegselmann and Ulrich Krause.
\newblock Opinion dynamics and bounded confidence--models, analysis, and
  simulation.
\newblock {\em J. Art. Soc. Soc. Sim.}, 5(3):paper 2, 2002.

\bibitem{Heitmann2006}
Katrin Heitmann, David Higdon, Charles Nakhleh, and Salman Habib.
\newblock Cosmic calibration.
\newblock {\em The Astrophysical Journal Letters}, 646(1):L1, 2006.

\bibitem{Higdon2005}
Dave Higdon, Marc Kennedy, James~C. Cavendish, John~A. Cafeo, and Robert~D.
  Ryne.
\newblock Combining field data and computer simulations for calibration and
  prediction.
\newblock {\em SIAM J. Sci. Comput.}, 26(2):448--466, February 2005.

\bibitem{Hirschman1970}
A.O. Hirschman.
\newblock {\em Exit, voice, and loyalty: Responses to decline in firms,
  organizations, and states}, volume~25.
\newblock Cambridge, Mass.: Harvard University Press, 1970.

\bibitem{Kendall2005a}
BE~Kendall, SP~Ellner, E~McCauley, SN~Wood, CJ~Briggs, WM~Murdoch, and
  P~Turchin.
\newblock Population cycles in the pine looper moth: {D}ynamical tests of
  mechanistic hypotheses.
\newblock {\em Ecological Monographs}, 75({2}):259--276, May 2005.

\bibitem{Kennedy2001}
Marc~C. Kennedy and Anthony O'Hagan.
\newblock Bayesian calibration of computer models.
\newblock {\em Journal of the Royal Statistical Society: Series B (Statistical
  Methodology)}, 63(3):425--464, 2001.

\bibitem{Klemm2003}
Konstantin Klemm, Víctor~M. Eguíluz, Raúl Toral, and Maxi~San Miguel.
\newblock Global culture: A noise-induced transition in finite systems.
\newblock {\em Phys. Rev. E}, 67:045101, Apr 2003.

\bibitem{Lazer2009}
David Lazer, Alex Pentland, Lada Adamic, Sinan Aral, Albert-László Barabási,
  Devon Brewer, Nicholas Christakis, Noshir Contractor, James Fowler, Myron
  Gutmann, Tony Jebara, Gary King, Michael Macy, Deb Roy, and Marshall
  Van~Alstyne.
\newblock Computational social science.
\newblock {\em Science}, 323(5915):721--723, 2009.

\bibitem{Lerner2002a}
Josh Lerner and Jean Tirole.
\newblock Some simple economics of open source.
\newblock {\em The Journal of Industrial Economics}, 50(2):197--234, 2002.

\bibitem{Malmgren2008a}
R~Dean Malmgren, Daniel~B Stouffer, Adilson~E Motterb, and Lu\'is A~N Amaral.
\newblock A poissonian explanation for heavy tails in e-mail communication.
\newblock {\em Proc. Natl. Acad. Sci. U. S. A.}, 105(47):18153--18158, Nov
  2008.

\bibitem{Maes2010}
Michael Mäs, Andreas Flache, and Dirk Helbing.
\newblock Individualization as driving force of clustering phenomena in humans.
\newblock {\em PLoS Computational Biology}, 6(10):e1000959, October 2010.

\bibitem{McFadden1989}
Daniel McFadden.
\newblock A method of simulated moments for estimation of discrete response
  models without numerical integration.
\newblock {\em Econometrica}, 57(5):995--1026, September 1989.

\bibitem{McPherson2001a}
Miller McPherson, Lynn Smith-Lovin, and James~M Cook.
\newblock Birds of a feather: Homophily in social networks.
\newblock {\em Annual Review of Sociology}, 27(1):415--444, 2001.

\bibitem{Morris1995}
Max~D. Morris and Toby~J. Mitchell.
\newblock Exploratory designs for computational experiments.
\newblock {\em Journal of Statistical Planning and Inference}, 43(3):381--402,
  1995.

\bibitem{Moscovici1969}
S.~Moscovici, E.~Lage, and M.~Naffrechoux.
\newblock Influence of a consistent minority on the responses of a majority in
  a color perception task.
\newblock {\em Sociometry}, 32(4):pp. 365--380, 1969.

\bibitem{Neff2012}
J.G. Neff, D.~Laniado, K.~Kappler, Y.~Volkovich, P.~Arag{\'o}n, and
  A.~Kaltenbrunner.
\newblock Jointly they edit: examining the impact of community identification
  on political interaction in wikipedia.
\newblock {\em arXiv preprint arXiv:1210.6883}, 2012.

\bibitem{Opp2001}
Karl-Dieter Opp.
\newblock How do norms emerge? an outline of a theory.
\newblock {\em Mind \& Society}, 2:101--128, 2001.
\newblock 10.1007/BF02512077.

\bibitem{Ostrom2000}
Elinor Ostrom.
\newblock Collective action and the evolution of social norms.
\newblock {\em The Journal of Economic Perspectives}, 14(3):pp. 137--158, 2000.

\bibitem{Parisi2003}
Domenico Parisi, Federico Cecconi, and Francesco Natale.
\newblock Cultural change in spatial environments.
\newblock {\em Journal of Conflict Resolution}, 47(2):163--179, 2003.

\bibitem{Radicchi2009a}
Filippo Radicchi.
\newblock Human activity in the web.
\newblock {\em Phys. Rev. E: Stat., Nonlinear, Soft Matter Phys.},
  80(2):026118, Aug 2009.

\bibitem{Rafaeli2008a}
Sheizaf Rafaeli and Yaron Ariel.
\newblock {\em Psychological aspects of cyberspace: Theory, research,
  applications}, chapter Online Motivational Factors: Incentives for
  Participation and Contribution in Wikipedia, pages 243--267.
\newblock Cambridge University Press, 2008.

\bibitem{Reagle2007}
Joseph M.~Jr. Reagle.
\newblock {\em Good Faith Collaboration--The Culture of Wikipedia}.
\newblock The MIT Press, 2007.

\bibitem{Ren2012}
Y.~Ren, R.~Kraut, S.~Kiesler, and P.~Resnick.
\newblock {\em Evidence-based social design: Mining the social sciences to
  build online communities}, chapter Encouraging commitment in online
  communities, pages 77--124.
\newblock MIT Press, Cambridge, MA, 2012.

\bibitem{Saltelli2004a}
Andrea Saltelli, Stefano Tarantola, Francesca Campolongo, and Marco Ratto.
\newblock {\em Sensitivity Analysis in Practice--A guide to Assessing
  Scientific Models}.
\newblock John Wiley \& Sons, Ltd., 2004.

\bibitem{Santner2003a}
T.J. Santner, B.~Williams, and W.~Notz.
\newblock {\em The Design and Analysis of Computer Experiments}.
\newblock Springer-Verlag, NY, 2003.

\bibitem{Schelling1971}
Thomas~C. Schelling.
\newblock Dynamic models of segregation.
\newblock {\em The Journal of Mathematical Sociology}, 1(2):143--186, 1971.

\bibitem{Smith1993}
A.~A. Smith.
\newblock Estimating nonlinear time-series models using simulated vector
  autoregressions.
\newblock {\em Journal of Applied Econometrics}, 8(S1):S63--S84, 1993.

\bibitem{Smith2008a}
Anthony A.~Jr. Smith.
\newblock indirect inference.
\newblock In Steven~N. Durlauf and Lawrence~E. Blume, editors, {\em The New
  Palgrave Dictionary of Economics}. Palgrave Macmillan, Basingstoke, 2008.

\bibitem{Sobkowicz2009a}
Pawel Sobkowicz.
\newblock Modelling opinion formation with physics tools: Call for closer link
  with reality.
\newblock {\em Journal Artificial Societies and Social Simulation}, 12(1):11,
  2009.

\bibitem{Tajfel1982}
H~Tajfel.
\newblock Social psychology of intergroup relations.
\newblock {\em Annual Review of Psychology}, 33(1):1--39, 1982.

\bibitem{Turner1989}
J.~C. Turner.
\newblock {\em Rediscovering the social group: A self-categorization theory}.
\newblock Blackwell Publishers, London, 1989.

\bibitem{Wilkinson2008a}
Dennis~M Wilkinson.
\newblock Strong regularities in online peer production.
\newblock In {\em Proceedings of the 9th ACM conference on Electronic
  commerce}, Chicago, Illinois USA, 2008.

\bibitem{Windrum2007}
Paul Windrum, Giorgio Fagiolo, and Alessio Moneta.
\newblock Empirical validation of agent-based models: Alternatives and
  prospects.
\newblock {\em Journal of Artificial Societies and Social Simulation}, 10(2):8,
  2007.

\bibitem{Wood2010a}
Simon~N. Wood.
\newblock Statistical inference for noisy nonlinear ecological dynamic systems.
\newblock {\em Nature}, 466(7310):1102--U113, August 26 2010.

\bibitem{Wu2010}
Ye~Wu, Changsong Zhou, Jinghua Xiao, JÃ¼rgen Kurths, and Hans~Joachim
  Schellnhuber.
\newblock Evidence for a bimodal distribution in human communication.
\newblock {\em Proceedings of the National Academy of Sciences},
  107(44):18803--18808, 2010.

\bibitem{Zhang2012}
D.~Zhang, K.~Prior, and M.~Levene.
\newblock How long do wikipedia editors keep active?
\newblock In {\em Proceedings of the 8th International Symposium on Wikis and
  Open Collaboration (Wikisym `12)}, 2012.

\end{thebibliography}

\end{document}